\documentclass[fleqn,usenatbib]{mnras}

\usepackage[T1]{fontenc}

\DeclareRobustCommand{\VAN}[3]{#2}
\let\VANthebibliography\thebibliography
\def\thebibliography{\DeclareRobustCommand{\VAN}[3]{##3}\VANthebibliography}

\usepackage{graphicx}	
\usepackage{amsmath}	
\usepackage{tabularx}

\title[Methods for triggering 1D supernova explosions]{Comparison of three methods for triggering core-collapse supernova explosions in spherical symmetry}

\author[L. Imasheva et al.]{
Liliya Imasheva,$^{1,2}$ 
 Hans-Thomas Janka,$^{1}$\thanks{E-mail: thj@mpa-garching.mpg.de}
and Achim Weiss$^{1,2}$
\\
$^{1}$Max-Planck Institut f\"ur Astrophysik, Karl-Schwarzschild-Str. 1, 85748 Garching, Germany\\
$^{2}$Ludwig-Maximilians-Universit\"at M\"unchen, Geschwister-Scholl-Platz 1, 80539 Munich, Germany\\
}

\date{Accepted XXX. Received YYY; in original form ZZZ}

\pubyear{2025}

\begin{document}
\label{firstpage}
\pagerange{\pageref{firstpage}--\pageref{lastpage}}
\maketitle

\begin{abstract}
Despite the three-dimensional nature of core-collapse supernovae (CCSNe), simulations in spherical symmetry (1D) play an important role to study large model sets for the progenitor-remnant connection, explosion properties, remnant masses, and CCSN nucleosynthesis. To trigger explosions in 1D, various numerical recipes have been applied, mostly with gross simplifications of the complex microphysics governing stellar core collapse, the formation of the compact remnant, and the mechanism of the explosion. Here we investigate the two most popular treatments, piston-driven and thermal-bomb explosions, in comparison to 1D explosions powered by a parametric neutrino engine in the P-HOTB code. For this comparison we calculate CCSNe for eight stars and evolution times up to $10^4$\,s, targeting the same progenitor-specific explosion energies as obtained by the neutrino-engine results. Otherwise we employ widely-used (``classic'') modelling assumptions, and alternatively to the standard contraction-expansion trajectory for pistons, we also test suitably selected Lagrangian mass shells adopted from the neutrino-driven explosions as ``special trajectories.'' Although the $^{56}$Ni production agrees within roughly a factor of two between the different explosion triggers, neither piston nor thermal bombs can reproduce the correlation of $^{56}$Ni yields and explosion energies found in neutrino-driven explosions. This shortcoming as well as the problem of massive fallback witnessed in classical piston models, which diminishes or extinguishes the ejected nickel, can be largely cured by the special trajectories. These and the choice of the explosion energies, however, make the modelling dependent on pre-existing neutrino-driven explosion results.
\end{abstract}




\section{Introduction}
\label{sec:intro}

The study of core-collapse supernovae (CCSNe) has captivated astronomers and physicists for decades \citep[for reviews on theoretical developments, see, e.g.,][]{Bethe1990,Woosley+2002,Woosley+2005,2007PhR...442...38J,Janka2012,Burrows+2013,Foglizzo+2015,Janka+2016,Mueller2016,2017hsn..book.1053F,Janka2017,2021Natur.589...29B}. These cataclysmic events mark the explosive deaths of massive stars, releasing an extraordinary amount of energy and dispersing heavy elements into the cosmos. Understanding the intricate processes involved in stellar core collapse and explosions is crucial for unraveling the mysteries of stellar evolution, the nucleosynthesis of chemical elements, and the formation of gaseous and compact remnants of CCSNe.

In recent years, significant progress has been made in modelling and simulating CCSNe using sophisticated computational techniques. Both, axisymmetric (2D) \citep[e.g.,][]{Janka+2012,2016ApJ...818..123B,Summa+2016,2016ApJ...822...61C,2017ApJ...843....2H,Radice+2017,2021Natur.589...29B} and ab-initio three-dimensional (3D) simulations \citep[e.g.,][]{2014ApJ...786...83T,Melson+2015a,Melson+2015b,2015ApJ...807L..31L,2016ARNPS..66..341J,Mueller+2017,2018ApJ...855L...3O,OConnor+2018,Summa+2018,2019MNRAS.484.3307M,2019ApJ...881...36G,Vartanyan+2019,Stockinger+2020,2020MNRAS.491.2715B,2021MNRAS.503.4942O,2021ApJ...915...28B,Burrows+2024,Nakamura+2024}, are essential for understanding the underlying physics of the explosion in great depth and details, carving out the processes that are responsible for the explosion. However, due to the enormous complexity of such simulations, even with modern supercomputers only a limited number of selected cases can be investigated and require appreciable computational resources.

While the true nature of CCSNe involves multi-dimensional hydrodynamic instabilities such as convection and energy-dependent three-flavor neutrino transport, spherical symmetry (1D) and various kinds of approximations for the triggering mechanism of the explosions, often sacrificing the treatment of neutrinos and the neutron-star physics, serve as a useful starting point. These simplifications permit time-dependent simulations in one dimension, i.e., just with a radius dependence, with much less computational demands. Therefore they have been widely employed, in particular, for determining CCSN nucleosynthesis \citep[for an overview, see, e.g.,][and references therein]{2017hsn..book.1753U} and CCSN light curves and spectra \citep[e.g.,][]{2016ApJ...821...38S,Goldberg+2019,Goldberg+2020,2020ApJ...902...95L,Teffs+2020a,Teffs+2020b,Dessart+2021a,Dessart+2021b,Curtis+2021,Dessart+2023}. Beyond such applications, approximate treatments also allow for systematic studies of large sets of CCSN calculations that are essential for a better understanding of the fundamental dependencies of stellar core collapse on crucial nuclear and particle physics, thus aiding the development of more comprehensive multi-dimensional models. Moreover, 1D simplifications also facilitate the advancement of our knowledge of stellar evolution, chemical element formation, and of the chemical enrichment history of galaxies \citep[e.g.,][]{1995ApJS...98..617T,2003Ap&SS.284..539M,2015ApJ...808..132H,2020ApJ...900..179K,2021MNRAS.506.4131W,2024ApJS..272...15R,2025MNRAS.536.2135J}. Similarly, deeper insights into the statistics and birth-mass distributions of resulting compact objects, i.e., neutron stars and black holes, gain from such studies \citep[e.g.,][]{2008ApJ...679..639Z,OConnor+2011,2012ApJ...757...69U,2015ApJ...801...90P,2016ApJ...818..124E,2016ApJ...821...38S,Ebinger+2019,Ebinger+2020,2020ApJ...890...51E,Woosley+2020,Boccioli+2024,2025A&A...695A.122U}. Thus they provide crucial information to refine models and simulations, contribute to our understanding of the Universe's chemical evolution, and pave the way for multi-messenger astronomy including gravitational-wave and neutrino measurements that offer a more complete picture of these extraordinary astrophysical events.

Explosions in 1D, however, have to be induced through artificial explosion triggers, i.e., numerical treatments that create sufficiently strong shock waves to blow up the progenitor stars.
These triggers, of which the most commonly used ones are known as thermal bombs, pistons, and parametric neutrino engines, are designed to initiate and control CCSN explosions for the purpose of studying specific aspects of the processes, phenomena, and observables mentioned above.

The thermal-bomb method was introduced by K.~Nomoto and collaborators \citep[e.g.,][]{1988A&A...196..141S, 1989A&A...210L...5H, 1990ApJ...349..222T,2001ApJ...555..880N,2006NuPhA.777..424N}. It was also used in more recent works \citep[e.g.,][]{2010ApJ...719.1445M,2019ApJ...886...47S,2022ApJ...937..116K}, 
and it is the explosion trigger implemented in the 1D-codes MESA \citep[][]{2015ApJS..220...15P} and SNEC \citep[][]{2015ApJ...814...63M}. The thermal bomb ``mechanism'' is also used for simulating hypernovae and low-energy explosions \citep[e.g.,][]{2006NuPhA.777..424N,2008ApJ...673.1014U,2010IAUS..265...34N}.
In hydrodynamic simulations, the thermal bomb can be implemented as a parametrized energy deposition term in the energy equation. By introducing this artificial release of thermal energy in a chosen volume (mass for a Lagrangian, and spatial domain for an Eulerian approach) and over a defined period of time, the aim, in the best case, is to more or less closely mimic the energy input expected by the true physical processes in the core of the CCSN. In our previous paper \citep{2023MNRAS.518.1818I} we investigated different setups for such thermal bombs and came up with some general recommendations for suitable treatments and parameter choices, which will be mentioned in Section~\ref{bomb}.

The piston-driven explosion treatment was introduced by S.~Woosley and collaborators, see \cite{1988ApJ...330..218W,1995ApJS..101..181W,Woosley+2002,2007PhR...442..269W,2008ApJ...679..639Z}. The piston-driven explosion technique involves the injection of kinetic energy by a moving Lagrangian inner grid boundary as the numerical recipe for driving the CCSN explosion, essentially by accelerating the innermost matter on the computational grid outward with a rapid and strong push. This method shall mimic the impact of the outward going shock wave produced by the core bounce after the preceding core infall, which stops abruptly when the neutron star begins to form. Such an approach assumes that the stellar material exterior to the mass cut, which separates the initial compact remnant from the ejecta, receives mostly kinetic energy.

Thermal bombs as well as pistons do not include neutrinos in their descriptions, which has consequences for the dynamics and the nucleosynthesis conditions in the CCSN models. In view of these limitations, the more sophisticated 1D neutrino engines to drive explosions by neutrino-energy deposition behind the CCSN shock in 1D hydrodynamic simulations were more recently introduced. These engines exist in various versions, which use different ingredients to achieve successful blast waves: enhanced postshock heating by electron neutrinos and antineutrinos ($\nu_e$ and $\bar\nu_e$) via a parametric increase of their absorption rates on neutrons and protons \citep{OConnor+2011,2025MNRAS.536.2135J}; a simple spherical and grey approximation of neutrino transport and increased $\nu_e$ and $\bar\nu_e$ luminosities from the proto-neutron star's high-density core \citep[P-HOTB;][]{2012ApJ...757...69U,2016ApJ...818..124E,2016ApJ...821...38S}; a neutrino-driven wind, which is injected at the inner grid boundary and blows out the overlying layers of the progenitor \citep{2015ApJ...801...90P}; energy-dependent, three-flavor, general-relativistic (GR) Boltzmann neutrino transport with a parameterization of additional postshock heating linked to the heavy-lepton neutrino ($\nu_\mu$, $\nu_\tau$, $\bar\nu_\mu$, $\bar\nu_\tau$) luminosities \citep[PUSH;][]{2015ApJ...806..275P}; or a 1D approximation of turbulent convection in the neutrino-heated postshock layer to aid explosions in 1D  Newtonian and GR neutrino-hydrodynamic models \citep[STIR;][]{2020ApJ...890..127C,Boccioli+2021,2022ApJ...934...67B}. All of these neutrino-engine methods were also used in large sets of 1D explosion models. In our present work we obtain neutrino-driven explosions in 1D CCSN simulations by the P-HOTB engine.

All these 1D explosion ``mechanisms'' require detailed designs of the additional energy inputs, which involve the definition of adjustable parameters. The main focus of the present study will be the question how these aspects influence the final outcomes of the 1D explosion calculations. For the final results of interest, we consider the dynamics of the developing CCSN explosions as well as their production of $^{56}$Ni and the masses of their compact remnants after the fallback of matter that does not get unbound during the first few hours. These aspects provide a suitable diagnostic basis to identify fundamental weaknesses and uncertainties \citep[e.g.][]{2009MNRAS.394.1317M,2015MNRAS.451..282S,2019MNRAS.483.3607S}, since especially the ejected mass of radioactive $^{56}$Ni has a pivotal importance for the light-curve modelling and can be directly inferred from observations \citep[e.g.][]{1989ARA&A..27..629A,1994ApJ...437L.115I}. 

Several studies have already been conducted in the past to compare different triggers of 1D explosions. \citet{1991ApJ...370..630A} considered thermal bomb and piston descriptions for the cases of collapsed and un-collapsed stellar cores before the initiation of the explosions. 
\citet{2007ApJ...664.1033Y} focused their investigation on obtaining different values of the observable final energies (from $0.8$ to $2.0\cdot 10^{51}$\,erg). They included neutrino effects during the collapse phase of the explosion, but switched them off afterwards. Their main concern were the remnant masses, which they considered in the context of investigating fallback and black-hole formation. 
In our work presented here we want to extend the research by \citet{1991ApJ...370..630A} by comparing thermal-bomb and piston-triggered explosions to our neutrino-engine-driven explosions. We will show that some of the results published from previous 1D simulations depend sensitively on the chosen numerical setups. 

In Section~\ref{sec:methods} we provide a concise overview of the pre-collapse stellar evolution models, the methodology employed for our hydrodynamic explosion modelling, the setups used for pistons, thermal bombs, and neutrino engines, and the small nuclear reaction network implemented in our hydrodynamic code. In Section~\ref{sec:results}, we present the results of our investigations, followed by a discussion in Section~\ref{sec:discussion} and by our conclusions in Section~\ref{sec:conclusion}.


\section{Methods and inputs}
\label{sec:methods}

In this work we investigate three different methods, which we will also call ``mechanisms,'' to trigger CCSN explosions in 1D simulations. We apply these trigger mechanisms to 1D pre-supernova stellar evolution models from \citet{2014ApJ...783...10S}, which were also employed in our previous paper \citep{2023MNRAS.518.1818I}. Our explosion simulations are performed with the \textsc{Prometheus-HOTB} hydrodynamics code, which was developed at the Max Planck Institute for Astrophysics to simulate CCSN explosions in 1D, 2D, and 3D with parametric neutrino engines \citep[P-HOTB;][]{1996A&A...306..167J,2003A&A...408..621K,2006A&A...457..963S,2007A&A...467.1227A,2012ApJ...757...69U,2016ApJ...818..124E}. The P-HOTB engine was used in 1D simulations to study the progenitor-explosion-remnant connections for very large sets of red supergiant and helium stars \citep{2016ApJ...821...38S,2016ApJ...818..124E,2020ApJ...890...51E}. The \textsc{Prometheus-HOTB} code was also supplemented with piston and thermal-bomb explosion triggers. In the following we describe the main aspects of the numerical methods employed, in particular the basic ingredients of the three explosion methods compared in our study, namely the neutrino-engine, piston, and thermal-bomb mechanisms.

\subsection{Presupernova models}
\label{sec:psn}

\begin{table}
	\centering
	\caption{Properties of the progenitors used in this work. $M_{\rm pre}$ is the total pre-collapse mass, $M_{\rm He}$ is the mass of the helium core, $M_{\rm CO}$ the mass of CO core, $M_{s=4}$ is the enclosed mass where the dimensionless entropy reaches the value $s/k_\mathrm{B} = 4$, and $M_{\rm Fe}$ is the iron core mass; see the definitions in Section~\ref{sec:psn}. All masses are in $M_\odot$.}
	\label{tab:psn}
  \begin{tabular}{ c|c|c|c|c|c  }
  \hline
  \hline
    $M_{\rm ZAMS}$ & $M_{\rm pre}$ & $M_{\rm He}$ & $M_{\rm CO}$ & $M_{\rm s=4}$ & $M_{\rm Fe}$ \\ 
    \hline
      13.6   &   11.7666  &    3.83755   &   2.66724   &   1.65399   & 1.43092 \\ 
      14.5   &   12.2015  &    4.20968   &   3.00563   &   1.80985   &  1.45020 \\ 
      15.1   &   12.9264  &    4.33190   &   3.14234   &   1.43694   &  1.42487 \\ 
      16.2   &   13.5441  &    4.77075   &   3.55961   &   1.51282   &  1.37969 \\ 
      19.8   &   15.8430  &    6.12564   &   4.88863   &   1.60033   &  1.44628 \\ 
      21.0   &   16.1109  &    6.62284   &   5.37384   &   1.48435   &  1.43162 \\ 
      21.7   &   16.3813  &    6.89419   &   5.63973   &   1.65120   &  1.46633 \\ 
      26.6   &   15.3093  &    8.96794   &   7.69495   &   1.73833   &  1.53154 \\ 
\hline 
\end{tabular} 
\end{table}

The presupernova models by \citet{2014ApJ...783...10S} were calculated using the 1D hydrodynamic stellar-evolution code KEPLER \citep{1978ApJ...225.1021W}. They represent non-rotating stars of solar metallicity, and the physical details were reported in an extended body of literature \citep[e.g.,][]{Woosley+2002, 2007PhR...442..269W}.

To decipher the differences between the three considered explosion mechanisms, we selected eight red supergiant progenitors with zero-age-main-sequence (ZAMS) masses of $M_{\rm ZAMS} = 13.6$, 14.5, 15.1, 16.2, 19.8, 21.0, 21.7, and $26.6\,M_\odot$. These progenitors were chosen to cover a wide range in the ZAMS mass as well as in the final explosion energies and remnant masses at the end of the neutrino-driven explosion simulations, which will be discussed in Section~\ref{sec:1d}. Their properties are summarized in Table~\ref{tab:psn}, where $M_{\rm pre}$ is the total pre-collapse mass, $M_{\rm He}$ is the helium-core mass defined by the mass coordinate where $X({\rm H})\le0.2$, $M_{\rm CO}$ is the mass of the carbon-oxygen core associated with the location where $X({\rm He})\le0.2$, $M_{s=4}$ is the mass enclosed by the radius where the value of the dimensionless entropy per nucleon is $s/k_\mathrm{B}=4$ (where $k_\mathrm{B}$ is Boltzmann's constant), and $M_{\rm Fe}$ is the iron core mass, defined as the point where the electron fraction $Y_e$ reaches $0.495$. 

\begin{figure*}
\includegraphics[width=\columnwidth]{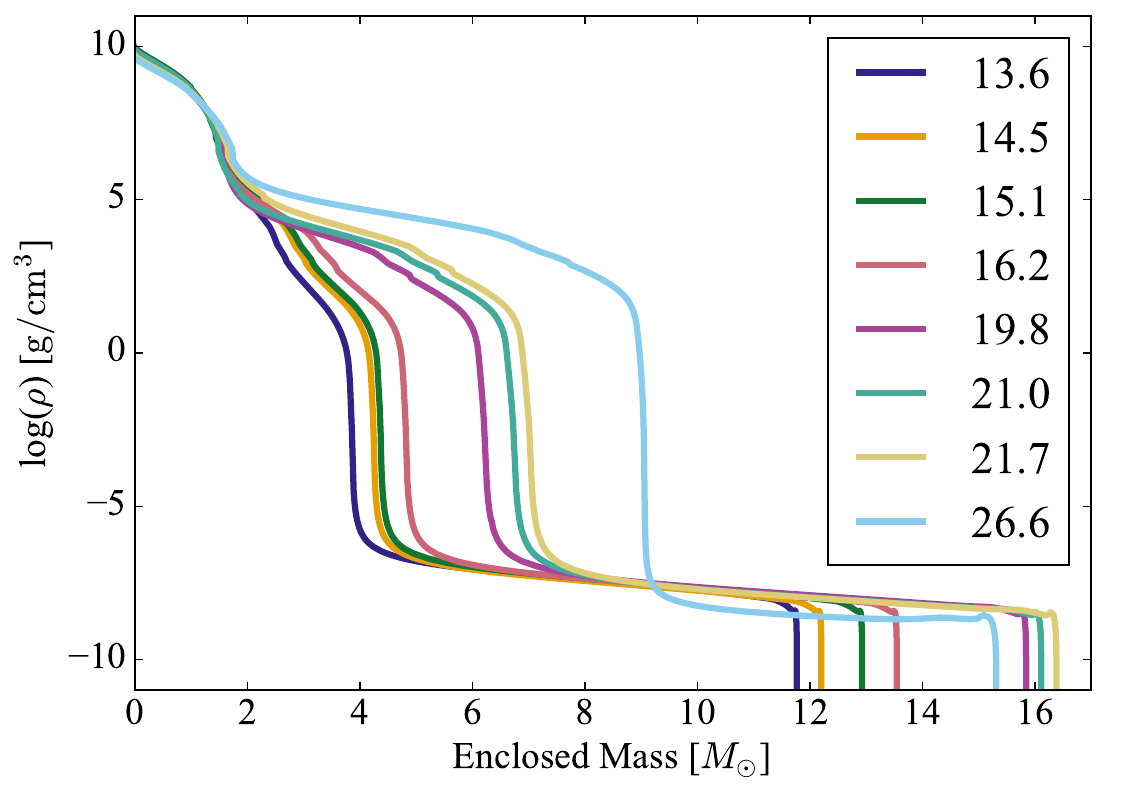}
\includegraphics[width=\columnwidth]{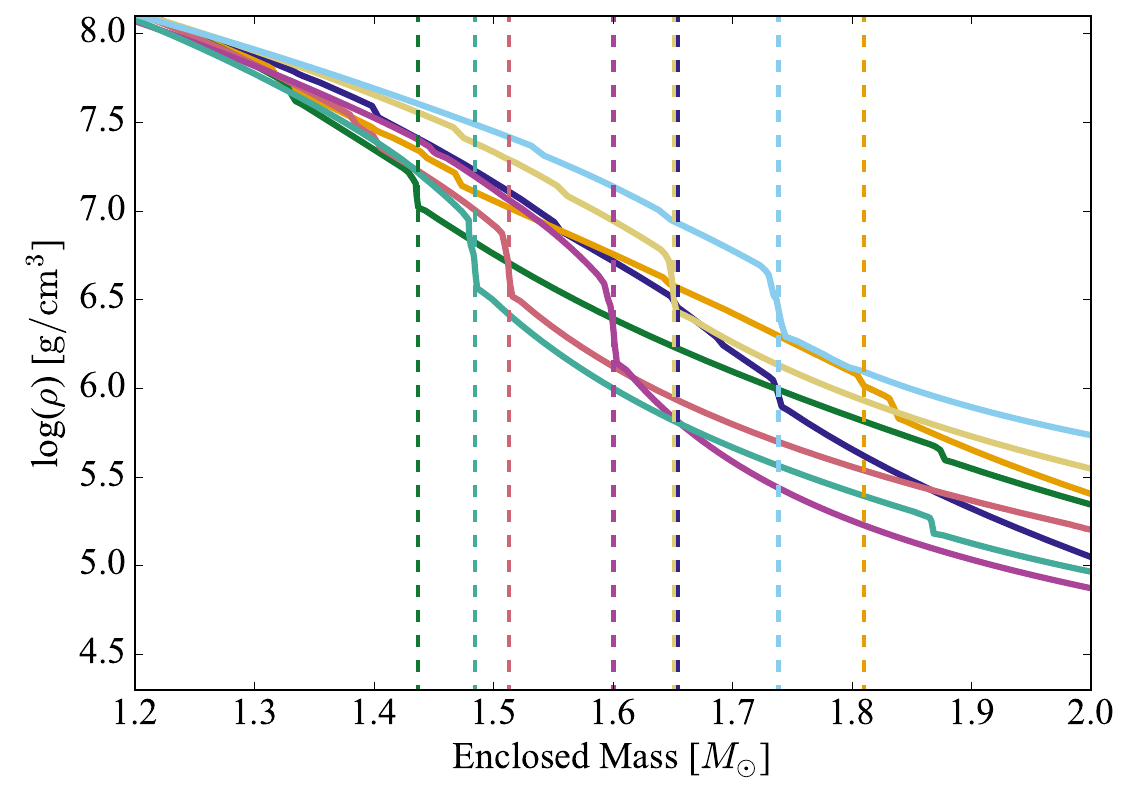}
\includegraphics[width=\columnwidth]{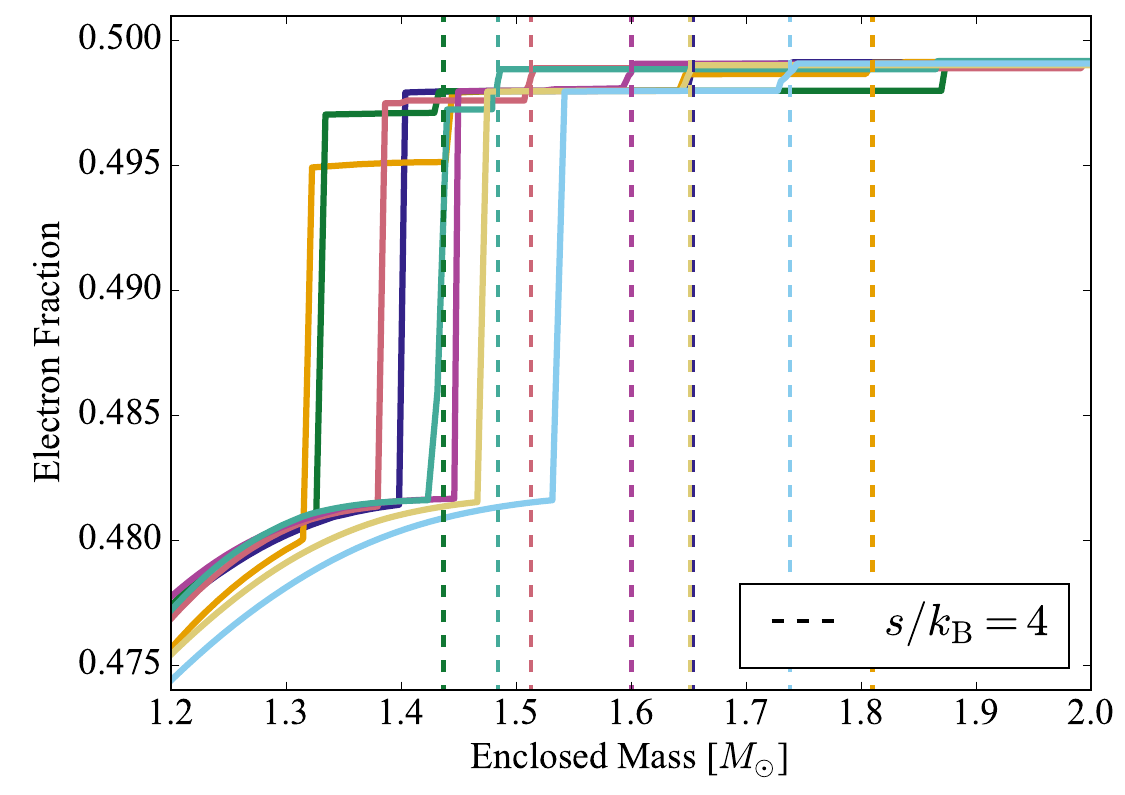}
\includegraphics[width=\columnwidth]{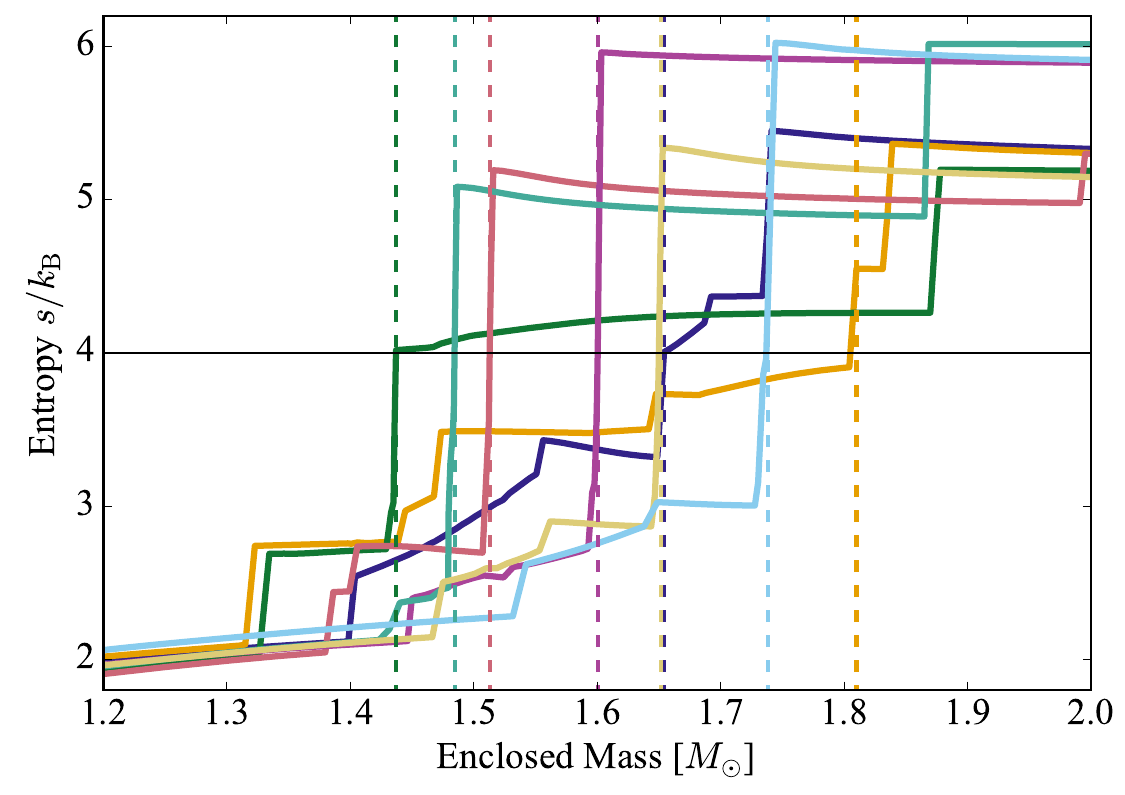}
\caption{Pre-collapse structure of the progenitor models used in this work: density (top left for the entire profiles and top right for the interval $[1.2,2.0]\,M_\odot$), electron fraction $Y_e$ (bottom left), and dimensionless entropy per nucleon $s/k_\mathrm{B}$ (bottom right) versus enclosed mass. Vertical dashed lines indicate the inner grid boundaries chosen in our explosion simulations for thermal-bomb and piston mechanisms; line colors correspond to the corresponding progenitors as listed in the inset. These lines are located at the points where the dimensionless entropy per nucleon $s/k_\mathrm{B}$ equals 4, which can also be seen by the horizontal black line in the $s/k_\mathrm{B}$ plot. 
}
\label{fig:psn_closer}
\end{figure*}

Figure~\ref{fig:psn_closer} shows the structure of all progenitors: density, electron fraction $Y_e$, and dimensionless entropy per nucleon as functions of enclosed mass. The vertical lines correspond to the locations at the base of the oxygen shell where the dimensionless entropy per nucleon is $s/k_\mathrm{B}=4$. This point will be used as inner grid boundary and thus as initial mass cut in our simulations of thermal-bomb and piston explosions.

\subsection{1D explosion modelling}
\label{sec:1d}

As already introduced in Section~\ref{sec:intro}, we are going to investigate three different ways to drive CCSNe explosions in 1D, namely by a:
\begin{enumerate}
    \item neutrino engine, using the P-HOTB method,
    \item piston,
    \item thermal bomb.
\end{enumerate}
In this section we describe the main ideas of these artificial explosion triggers and the details of their setups. All three explosion mechanisms have been implemented into the \textsc{Prometheus-HOTB} code. The code includes a high-density EoS to treat hot proto-neutron-star (PNS) matter as well as a low-density equation of state to deal with the conditions in the progenitor from the pre-collapse iron-core out to the surface of the hydrogen envelope.

In contrast to the other two mechanisms, the neutrino engine includes the PNS and handles its neutino emission and the neutrino physics that leads to shock revival. The initial PNS mass and the explosion energy are no free parameters in this case, but they are outcomes of the simulations once the neutrino engine is specified and values for its free parameters have been chosen, e.g., via calibration with suitable observations or results of ab-initio 3D CCSN models. The details of this explosion mechanism will follow below in Sect.~\ref{sec:neutrino}.

Comparing the three mechanisms with each other is therefore by no means straightforward. We did it such that their respective setups were tuned to yield the same explosion energy as the neutrino-driven explosion for the same progenitor, which we could achieve within 3\% accuracy in the worst cases. The explosion energy is defined here as the integral of the sum of the kinetic, internal and gravitational energies for all mass shells in the postshock domain with a positive binding energy, i.e., a positive total of these three energies. The terminal value of the explosion energy is usually reached after around $80$\,s of simulation time for thermal bomb and piston-driven explosions.

\subsubsection{Neutrino-driven explosions}
\label{sec:neutrino}


The neutrino engine to obtain neutrino-driven explosions in 1D consists of a two-stage treatment from the onset of the collapse of the stellar core until the neutrino-driven wind becomes dynamically unimportant, from where on the third phase is the purely hydrodynamic long-term evolution, which can be calculated without including neutrino effects. Here we just summarize the main aspects of the method. For more details, see \cite{2016ApJ...818..124E} and \cite{2016ApJ...821...38S}. 

During the collapse phase the simulations are carried on until core bounce at central densities exceeding the saturation density of nuclear matter ($\sim2.5\cdot 10^{14}$\,g/cm$^3$). This allows us to self-consistently catch the formation of the CCSN shock. The deleptonization (by electron captures and $\nu_e$ loss) during the collapse phase is followed by a simple and efficient numerical scheme proposed by \cite{2005ApJ...633.1042L}. It implies a replacement of the numerically more demanding solution of a transport problem by a local problem, which approximates the $Y_e$ evolution in the iron core by a predefined function of the matter density $\rho$.
The definition of the $Y_e(\rho)$ behaviour is based on results from core-collapse simulations with elaborate neutrino transport.

The infall of the subsonically collapsing inner core abruptly stops at bounce, and the re-expansion of this central body into the supersonically collapsing outer layers of the iron core leads to the formation of a shock wave that begins to expand outward. At this point, starting from shortly after core bounce, it is crucial to follow the neutrino production and propagation, which our neutrino engine handles by integrating the grey (i.e., energy-integrated) neutrino transport equation in time and radius in a numerically fast analytic manner, which is described in detail in \cite{2006A&A...457..963S}. This solver implies that instead of the full neutrino energy spectra of $\nu_e$, $\bar\nu_e$, and heavy-lepton neutrinos only the luminosities and number fluxes of these neutrino species are evolved in space and time. 

During this second phase, which is continued until $\sim$10\,s after core bounce, the neutrino cooling of the PNS has to be calculated in order to obtain the time evolution of the neutrino luminosities and mean energies that are needed to determine the neutrino energy deposition that initiates and powers the neutrino-driven explosion. Such calculations require complex codes, whose application is computationally quite expensive. Therefore, instead of taking this route, P-HOTB employs an analytic one-zone model for the Kelvin-Helmholtz neutrino cooling and concomitant contraction of the high-density core of the PNS. In practice, the central region of the collapsed stellar core is excluded from the computational grid only a few milliseconds after core bounce and is replaced by a retreating inner grid boundary mimicking the contraction of the PNS's high-density core. The one-zone core model provides the boundary conditions at the inner boundary of the computational domain. The rest of the collapsing star including the outer layers of the PNS, typically up to densities of at most several $10^{13}$\,g\,cm$^{-3}$ and neutrino mean optical depths of a few 100, are continued to be hydrodynamically tracked, using the grey neutrino treatment mentioned above. 

The one-zone core is considered to lose energy by radiating neutrinos with a total luminosity, $L_{\nu,{\rm c}}(t)$, whose time dependence can be parametrized by employing total energy conservation and the virial theorem \citep[for details, see][]{2012ApJ...757...69U,ertl_2016}. The total core-neutrino luminosity is coined as a function of the constant core mass, $M_{\rm c}$, the core radius, $R_{\rm c}(t)$, the rate of contraction of this radius, $\dot R_{\rm c}(t)$, the mass of the accretion mantle of the PNS around the core, $m_{\rm acc}$, (taken to be the mass between the inner grid boundary and a lower density of $10^{10}$\,g\,cm$^{-3}$), and the mass-accretion rate of the PNS, $\dot m_{\rm acc}$. The corresponding expression for the core luminosity can be written as
\begin{equation}
L_{\nu,{\rm c}}(t)=\frac{1}{3(\Gamma-1)}\Big[(3\Gamma-4)(E_{\rm g}+S)\,\frac{\dot R_{\rm c}}{R_{\rm c}}+S\,\frac{\dot m_{\rm acc}}{m_{\rm acc}}\Big]\,,
	\label{eq:neutrino}
\end{equation}
with the factors
\begin{eqnarray}
E_{\rm g}+S &=&-\frac{2}{5}\frac{GM_{\rm c}}{R_{\rm c}}\,\Big(M_{\rm c} +\frac52 \zeta m_{\rm acc}\Big)\,, 
	\label{eq:factor1} \\
S &=& -\zeta\, \frac{GM_{\rm c}m_{\rm acc}}{R_{\rm c}}\,,
	\label{eq:factor2}
\end{eqnarray}
where the mean adiabatic index $\Gamma$ of the nuclear matter in the high-density core and $\zeta$ ($0<\zeta\le1$) are free parameters of the approach.

The core radius as a function of time used in Equations~(\ref{eq:neutrino})--(\ref{eq:factor2}) is defined by
\begin{equation}
R_{\rm c}(t)=R_{\rm c,f}+\frac{R_{\rm c,i}-R_{\rm c,f}}{(1+t)^n}\,,
	\label{eq:neutrino_radius}
\end{equation}
where $R_{\rm c,i}$ is the initial core radius, which is set equal to the initial radius of the inner grid boundary, and $R_{\rm c,f}$ is the final core radius, which, along with the exponent, $n$, is also a free parameter of this approach.

To sum up, the high-density core of the PNS emerging from the iron-core collapse is excised from the computational volume and replaced by a contracting inner grid boundary shortly after the expanding shock has converted to a stalled accretion shock. The evolution of this central core is parametrically treated by the described one-zone model, which affects the hydrodynamics of the rest of the star by its contribution to the gravitational potential of the PNS and the total neutrino luminosity injected at the inner grid boundary, which is suitably distributed between the different neutrino species \citep[see][]{2012ApJ...757...69U}. 

A fixed combination of the parameters $\Gamma$, $\zeta$, $R_{\rm c}(t)$, and $n$ (see Equations~\ref{eq:factor1} to  \ref{eq:neutrino_radius}) defines the P-HOTB neutrino engine, and in this work we consider three different parameter sets (``calibration models'') for this neutrino engine, whose values are listed in Table~\ref{tab:neutrino}.

The neutrino engine has been calibrated in \citet{2016ApJ...821...38S} by referring to the observationally and theoretically constrained explosion energies and $^{56}$Ni masses of SN\,1987A and SN\,1054-Crab. The calibration models W18, S19.8, and W20 of Table~\ref{tab:neutrino} are taken from \cite{2016ApJ...821...38S}, where they are discussed in detail.
Model W18 results from a blue supergiant progenitor that produced a large amount of oxygen with enhancements in surface helium and nitrogen abundances. The total mass, helium-core mass, oxygen mass, and resulting PNS mass of model S19.8 are all in agreement with SN~1987A and blue supergiant models suggested for its progenitor Sk~$-69^\circ$\,202, but the S19.8 pre-supernova model was a red supergiant. Model W20 comes close to the observed luminosity and surface temperature of Sk~$-69^\circ$\,202 in the Large Magellanic Cloud (LMC) \citep{1987ApJ...321L..41W}. Further details about the calibration models can be found in \cite{2016ApJ...821...38S}.

During the third phase in the neutrino-driven explosion models, which starts at $\sim$10\,s after core bounce, neutrino effects are considered to be unimportant and the inner grid boundary for the subsequent long-term hydrodynamic evolution is moved to $10^9$\,cm and switched to an open (outflow) boundary.

\begin{table}
    \centering
        \caption{Parameter values of the PNS core model for the neutrino engine P-HOTB used in this work; see Equations~(\ref{eq:neutrino})--(\ref{eq:neutrino_radius}).}
    \begin{tabular}{ccccc}
    \hline
    \hline
Calibration model & $R_{\rm c,f}$ [km] & $\Gamma$ & $\zeta$ & $n$ \\ \hline
W18 &  $6.0$ & $3.0$ & $0.65$ & $3.06$ \\
S19.8 &  $6.5$ & $3.0$ & $0.90$ & $2.96$ \\
W20 &  $6.0$ & $3.0$ & $0.70$ & $2.84$ \\ 
\hline
    \end{tabular}
    \label{tab:neutrino}
\end{table}

\subsubsection{Piston-driven explosions}
\label{sec:piston}

The second mechanism is the piston-driven explosion trigger \citep{1982ena..conf..377W, 1986ARA&A..24..205W, 1988ApJ...330..218W,1995ApJS..101..181W,2001ApJ...550..410M,2016ApJ...821...38S}, where the shock wave is artificially generated by moving the inner (Lagrangian) boundary of the computational grid with a highly supersonic velocity. 
The inner grid boundary thus plays the role of the piston and of this initial location of the CCSN shock, and its time-dependent behaviour is governed by free parameters in this approach. The value of the enclosed mass, where the piston is placed, is called the ``piston mass,'' $M_{\rm pist}$. It is defined by the choice of a proper location for the inner grid boundary in the radial structure of the progenitor model, and it also termed the ``mass cut'', which defines the initial separation point between compact remnant and CCSN ejecta.

To mimic the collapse phase before the explosion, the piston is initially moved inward for a time period $t_{\rm coll}$ until it reaches a minimum radius $r_{\rm min}$. At this time the inward motion is stopped and superseded by an outward movement of the piston with a velocity $u_0$, which is the parameter controlling the explosion. The piston velocity as a function of time is given by \citep{1995ApJS..101..181W}:
\begin{equation}
\frac{dr}{dt}=\begin{cases}
  v_0 -a_0t &  t<t_{\rm coll}, \\ 
  \sqrt{u_0^2+2fGM_{\rm pist}(1/r-1/r_{\rm min})} & t\ge t_{\rm coll},r<r_{\rm max}, \\ 
  0 & t\ge t_{\rm coll},r\ge r_{\rm max},
\end{cases}
	\label{eq:piston}
\end{equation}
where $v_0$ is the initial velocity in the progenitor model at the location where the piston is placed; $a_0=2(r_0-r_{\rm min}+t_{\rm coll}v_0)/t_{\rm coll}^2$ is a constant acceleration calculated in order to reach the minimum radius, $r_{\rm min}$, within the time interval $t_{\rm coll}$, with $r_0$ being the initial piston radius; $f=-u^2_0/[2GM_{\rm pist}(1/r_{\rm max}-1/r_{\rm min})]$ is chosen in order to ensure that the piston coasts to an asymptotic radius of $r_{\rm max}$. The piston is then held at the maximum radius.

Equation~(\ref{eq:piston}) determines the time-dependent position of the inner grid boundary, because the piston is located at a constant Lagrangian mass coordinate. The collapse phase is controlled by the parameters $t_{\rm coll}$ and $r_{\rm min}$, whereas the explosion phase in controlled by the parameters $u_0$ and $r_{\rm max}$. The explosion velocity $u_0$ is given to the piston at $t_{\rm coll}$, leading to an expansion of the grid boundary and its surrounding mass shells.

The ``classical'' procedure of the piston method was introduced by S.~Woosley and collaborators \citep{1995ApJS..101..181W,2001ApJ...550..410M,2008ApJ...679..639Z}, where the parameters were set to $t_{\rm coll}=0.45$\,s and $r_{\rm min}=5\cdot10^7$\,cm. The mass cut was placed at the base of the oxygen shell (defined as the point where the dimensionless entropy $s/k_\mathrm{B}$ is equal to 4; see the fifth column in Table~\ref{tab:psn}). In the present work, the collapse phase is simulated by using this method with fixed $t_{\rm coll}$ and $r_{\rm min}$ for all progenitors. The inner boundary is kept closed and reflective for the first $100$\,s to provide the necessary pressure to the model while the explosion is still developing \citep{2001ApJ...550..410M}. After $100$\,s the boundary is open, which allows for the matter to fall back onto the PNS. The piston velocity $u_0$ is the parameter varied in order to get the desired explosion energy, which in our case is pre-defined by a neutrino-driven explosion model for each considered progenitor.

\begin{table}
	\centering
	\caption{Parameter values for the collapse phase with the special trajectories in our hydrodynamic simulations of piston-driven explosions, where $mM_{\rm ZAMS}$ represents the mechanism $m$ (``PS'' for piston with special trajectory) and the ZAMS mass, $t_{\rm coll}$ is the time of bounce, and $r_{\rm min}$ is the minimum radius and thus the location of the inner grid boundary at bounce.}
	\label{tab:piston_special_par}
  \begin{tabular}{ c|c|c }
  \hline
  \hline
    $mM_{\rm ZAMS}$ [$M_\odot$] & $t_{\rm coll}$ [s] & $r_{\rm min}$ [$10^7$\,cm] \\
 \hline
PS13.6  & 0.853 & 1.328588\\
PS14.5  & 1.271 & 1.322752\\
PS15.1  & 1.842 & 1.267855\\
PS16.2  & 1.070 & 1.283356\\
PS19.8  & 0.575 & 1.372842\\
PS21.0  & 1.040 & 1.298096\\
PS21.7  & 1.603 & 1.363798\\
PS26.6  & 1.094 & 1.233201\\
\hline 
\end{tabular} 
\end{table}

By changing the parameters $t_{\rm coll},\,r_{\rm min}$, and $M_{\rm pist}$ it is possible to set up a more physical trajectory of the inner boundary that can mostly avoid the massive fallback seen in simulations with the classic piston in many cases (see subsection \ref{sec:fallphase}). 
\cite{2016ApJ...821...38S} introduced this alternative prescription as the so-called ``special-trajectory'' approach, which we also adopt here for comparison. The idea of this improvement is to try to mimic the dynamics of the neutrino-driven explosions more closely in piston-driven models. These special trajectories are defined by the time-dependent radial positions of those Lagrangian mass shells that are the first to follow the outward moving shocks in the neutrino-driven explosion models. Using these trajectories changes the time $t_{\rm coll}$ and the location $r_{\rm min}$ of the bounce as well as the location of the initial mass cut.

Figure~\ref{fig:shells}, top panel, displays the radius evolution with time for selected Lagrangian mass shells in the neutrino-driven explosion simulation of the 21\,$M_\odot$ progenitor. The red line indicates the mass enclosed by the trajectory $M_{\rm special}$, which is identified as the first mass shell crossing the shock radius (blue line) when the shock is revived by neutrino heating and starts to move outward after its stagnation. Superimposed on the mass-shell evolution of the neutrino-driven explosion model, the green line shows the movement of the inner grid boundary for the piston as constructed by the special-trajectory method with $M_{\rm pist} = M_{\rm special}$. Obviously, this trajectory reproduces important features of the special trajectory $M_{\rm special}$ of the neutrino-driven explosion; both have similar collapse times and similar minimum radii reached in the infall phase.

In this alternative prescription, the parameters $t_{\rm coll}$, $r_{\rm min}$, and the location of the mass cut are defined by the corresponding neutrino-driven explosion, and they are thus specific for each individual progenitor. The values of  $t_{\rm coll}$, $r_{\rm min}$ for all progenitors can be found in Table~\ref{tab:piston_special_par}. The explosion velocity $u_0$ has to be adjusted, too, in order to get the desired explosion energy. The latter is also taken from the neutrino-driven explosion model of each progenitor. The special-trajectory piston models therefore need the corresponding neutrino-driven models to find the optimal parameter values.

\subsubsection{Thermal-bomb explosions}
\label{bomb}

Finally, in the thermal-bomb approach \citep{1991ApJ...370..630A,2007ApJ...664.1033Y,2019ApJ...886...47S} the explosion is triggered by an instantaneous or a longer-lasting energy input in a spatial volume or mass interval beginning at the inner boundary (IB) of the computational grid. The inner boundary thus represents the initial PNS surface, and its location (again called mass cut) is the first parameter of this method. The PNS itself is again removed from the simulation region and replaced by the inner boundary. Additional free parameters of the model are the injected (or deposited) energy, $E_{\rm inj}$, the volume (by mass in our case) of this deposition, $\Delta M$, and the time during which the energy is deposited, $t_{\rm inj}$. This procedure provides more flexibility compared to piston-driven explosions, where the explosion sets in instantaneously. The mentioned parameters define the specific rate of energy input, $E_{\rm inj} /(t_{\rm inj}\,\Delta M)$. Being constant, this results in a linear increase of the energy input, which pushes the mass shells outward, leading to the explosion. The thermal-bomb mechanism is rather adaptable; it is easy to use, and the variation of the parameters can control the dynamics of the explosion. In \citet{2023MNRAS.518.1818I} we investigated different parametrizations of this mechanism and came up with general recommendations for modeling thermal-bomb explosions, in particular to collapse the energy-deposition shells before injecting the explosion-triggering energy. These recommended settings will be used for our thermal-bomb models in the present work.

For the thermal-bomb approach, the same collapse treatment as in the classic piston is applied. To achieve the same explosion energy as for the neutrino-driven explosions, the total injected energy $E_{\rm inj}$ has to be adjusted. It is deposited in the innermost $0.05 M_\odot$ of the computational grid, i.e., neighbouring the outer edge of the PNS. The energy-growth timescale is taken to be $1.0$\,s. The initial mass cut (inner grid boundary) is placed at the base of the oxygen shell, which is defined by the entropy value of $s/k_\mathrm{B} = 4$ as before. 

For all the thermal-bomb simulations, a reflecting (closed) inner boundary condition is employed for the first $10$\,s in order to mimic the neutrino-driven simulations. Then it is switched to an outflow (open) boundary.

\subsection{Reaction network}
\label{sec:networks}

As discussed in \cite{2023MNRAS.518.1818I}, the \textsc{Prometheus-HOTB} code has a small $\alpha$-capture reaction network that is meant to keep track of the bulk nucleosynthesis \citep{1986A&A...162..103M}. The isotopes included in the network are $^{4}$He, $^{12}$C, $^{16}$O, $^{20}$Ne, $^{24}$Mg, $^{28}$Si, $^{32}$S, $^{36}$Ar, $^{40}$Ca, $^{44}$Ti, $^{48}$Cr, $^{52}$Fe, and $^{56}$Ni, plus a ``tracer nucleus'' $^{56}$Tr, which keeps track of the formation of neutron-rich isotopes in matter with considerable neutron excess, i.e., when $Y_e < 0.49$ \citep{2000ApJ...531L.123K,2001AIPC..561...21K,2003A&A...408..621K}. The network is coupled to the hydrodynamics calculations, so the nuclear energy generation is taken into account. The reaction rates for the network are adopted from \citet{1996ApJ...460..408T}. Explicit network calculations are performed for temperatures between $0.1$\,GK and $9$\,GK. For even higher temperatures nuclear statistical equilibrium (NSE) is assumed. The network is not meant to calculate the detailed nucleosynthesis, but it is able to provide a reliable result for the energy release in nuclear reactions, especially during the explosive nuclear burning. Also the final masses of iron-group and alpha elements, as well as their evolution with time, are reasonable well represented.

\section{Results}
\label{sec:results}

In this section we present the results of our simulations for all three explosion mechanisms, focusing on the dynamics of the explosions, fallback masses, compact-remnant masses, and the final nickel masses. Model names used is this paper are denoted by $mM_{\rm ZAMS}$, where $m$ represents the explosion mechanism and $M_{\rm ZAMS}$ gives the ZAMS mass of the progenitor. The letter $m$ stands for ``N'' in the case of neutrino-driven explosion models, ``PC'' for classic piston-driven explosions, ``PS'' for special-trajectory piston-driven explosions, and ``T'' for thermal-bomb triggered explosions. 

\subsection{Dynamical Evolution}
\label{sec:dynevo}

The dynamical behaviour of the models is affected by the mechanism used. This applies both for the early explosion development and for the long-term evolution. In this section we will take a closer look at how the treatment of the explosion can determine the outcome.

\subsubsection{Explosion phase}
\label{sec:explphase}

\begin{figure}
\begin{center}
	\includegraphics[width=0.99\columnwidth]{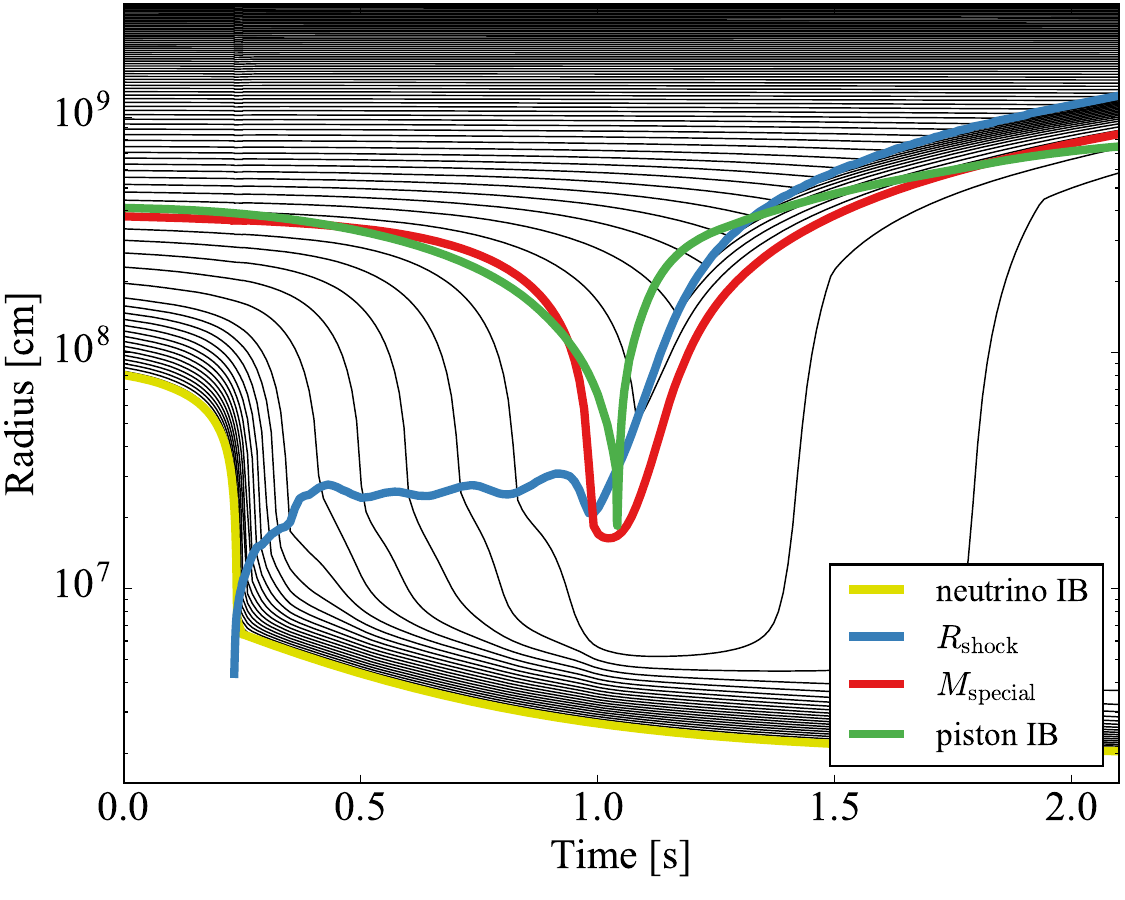}
	\includegraphics[width=0.99\columnwidth]{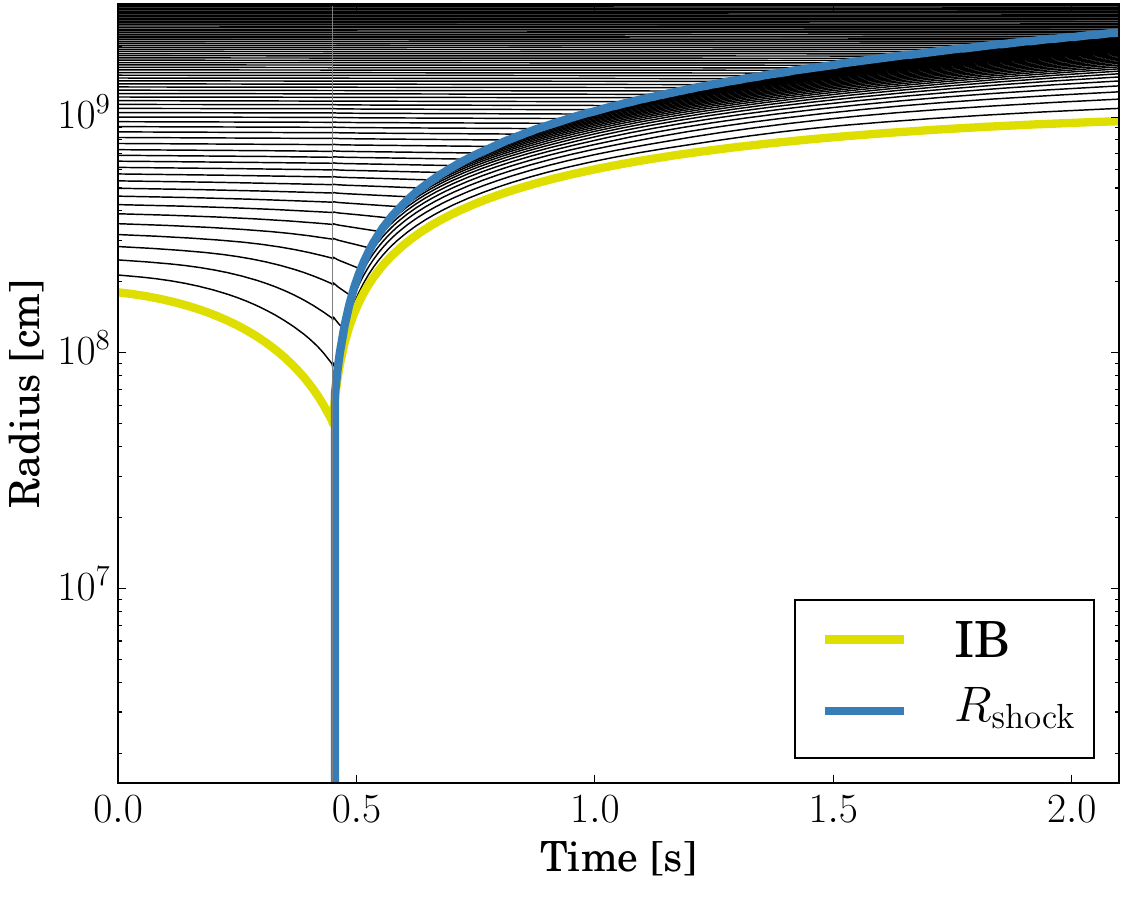}
	\includegraphics[width=0.99\columnwidth]{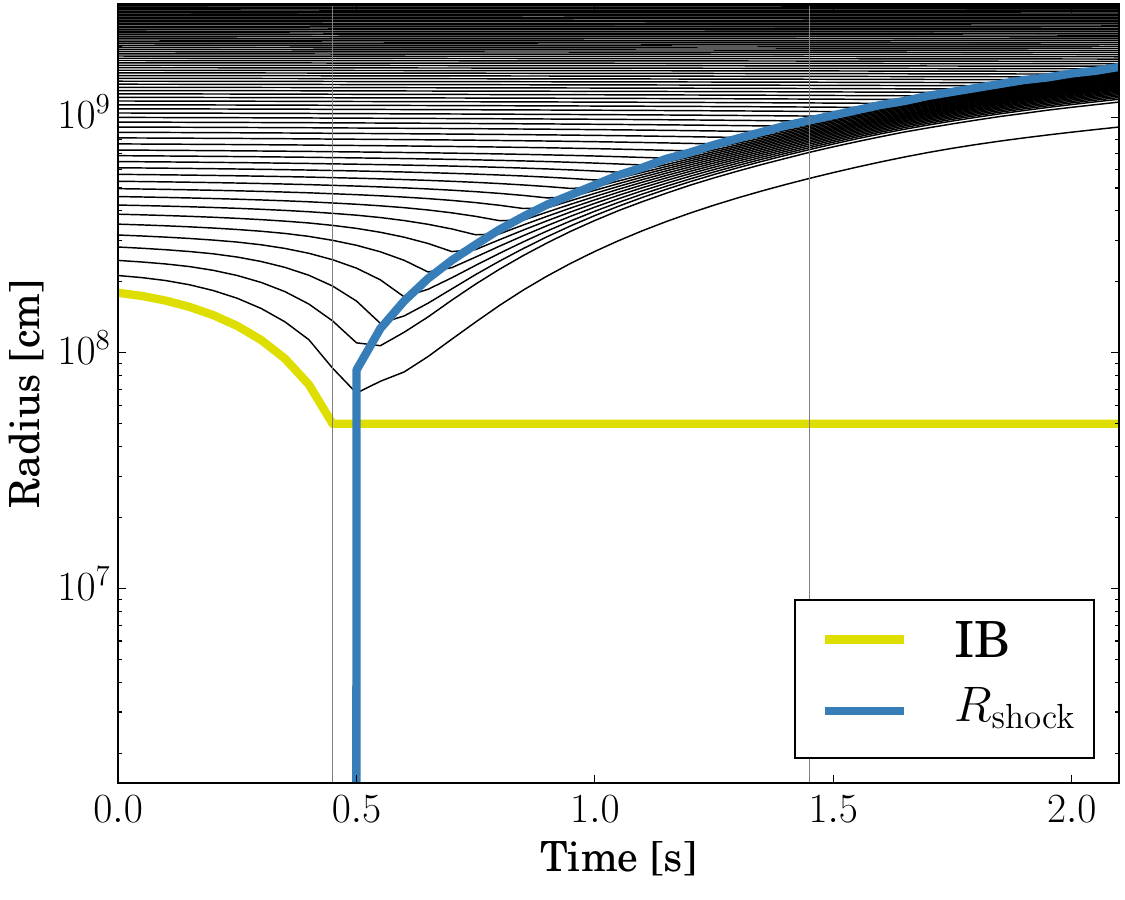}
 \end{center}
 \vspace{-15pt}
    \caption{Radius evolution of Lagrangian mass shells with time (measured from the start of the simulations) for the CCSN runs of the 21.0\,$M_\odot$ progenitor with different explosion mechanisms: neutrino-driven (top panel), classic piston-driven with a vertical line corresponding to the time of bounce at 0.45\,s (middle panel), and thermal bomb with vertical lines corresponding to the beginning and end of the energy deposition, which lasts for 1\,s in this case (bottom panel). The thin black solid lines are the mass shells, spaced in steps of 0.025\,$M_\odot$, the blue line marks the shock radius, and the yellow line represents the movement of the inner grid boundary.}
    \label{fig:shells}
\end{figure}

We first investigate how the explosion develops in the first couple of seconds after the stellar core begins to collapse. Figure~\ref{fig:shells} contains the mass shell plots for the neutrino-driven (top), piston (middle), and thermal bomb (bottom) mechanisms, applied to the same progenitor of $M=21.0\,M_\odot$. The red and green lines in the top panel have no direct meaning for the neutrino-driven explosions, but are relevant for setting up different piston models, and will be discussed later. In the neutrino-driven explosion model the shock (blue line) is formed by the core bounce at about 0.25\,s after the start of the simulation. After a short initial phase of expansion in mass and radius, the shock stalls by the interaction with the infalling matter till around $1.0$\,s. At this time a rapid outward acceleration of the shock takes place, initiated by neutrino heating in the gain layer behind the shock. Some of the matter that has transiently accumulated on the newly formed PNS is blown outward in a neutrino-driven wind as well. 

\begin{figure}
\begin{center}
	\includegraphics[width=0.97\columnwidth]{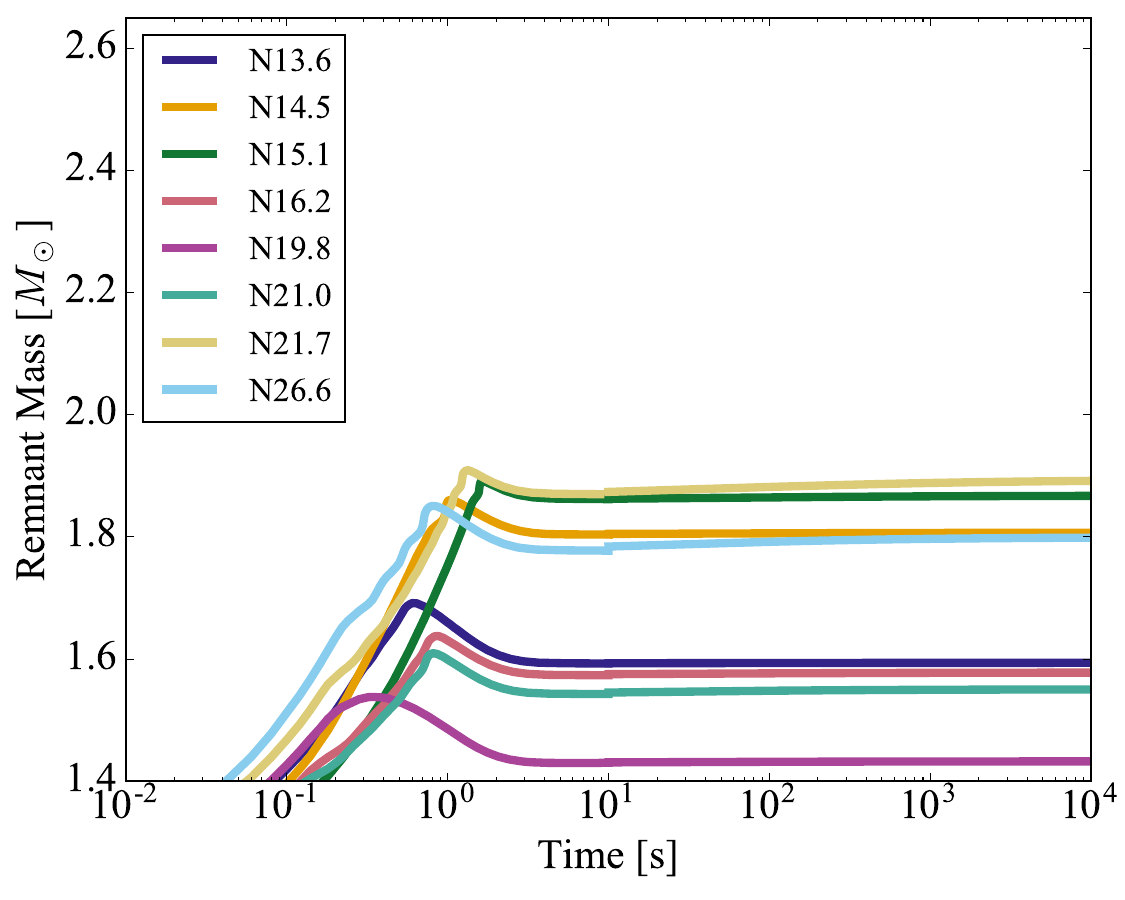}
	\includegraphics[width=0.97\columnwidth]{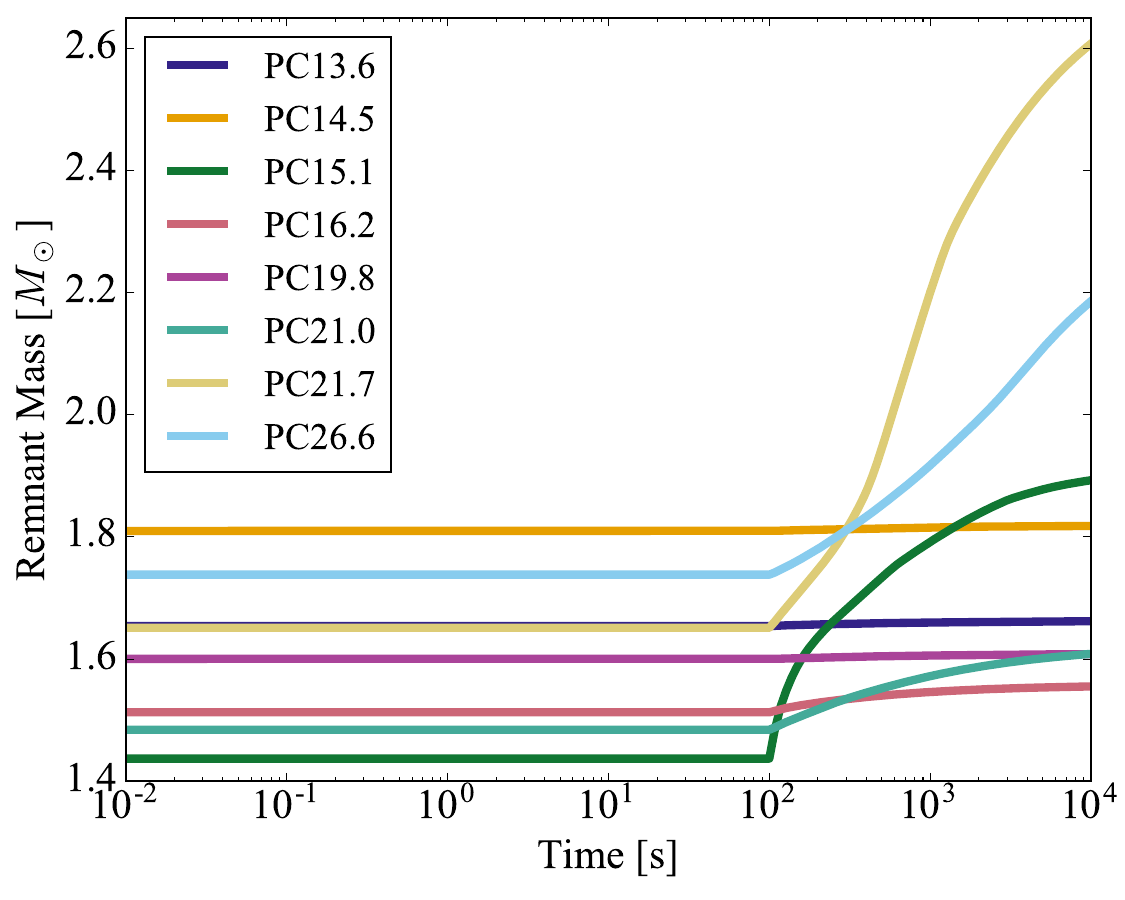}
	\includegraphics[width=0.97\columnwidth]{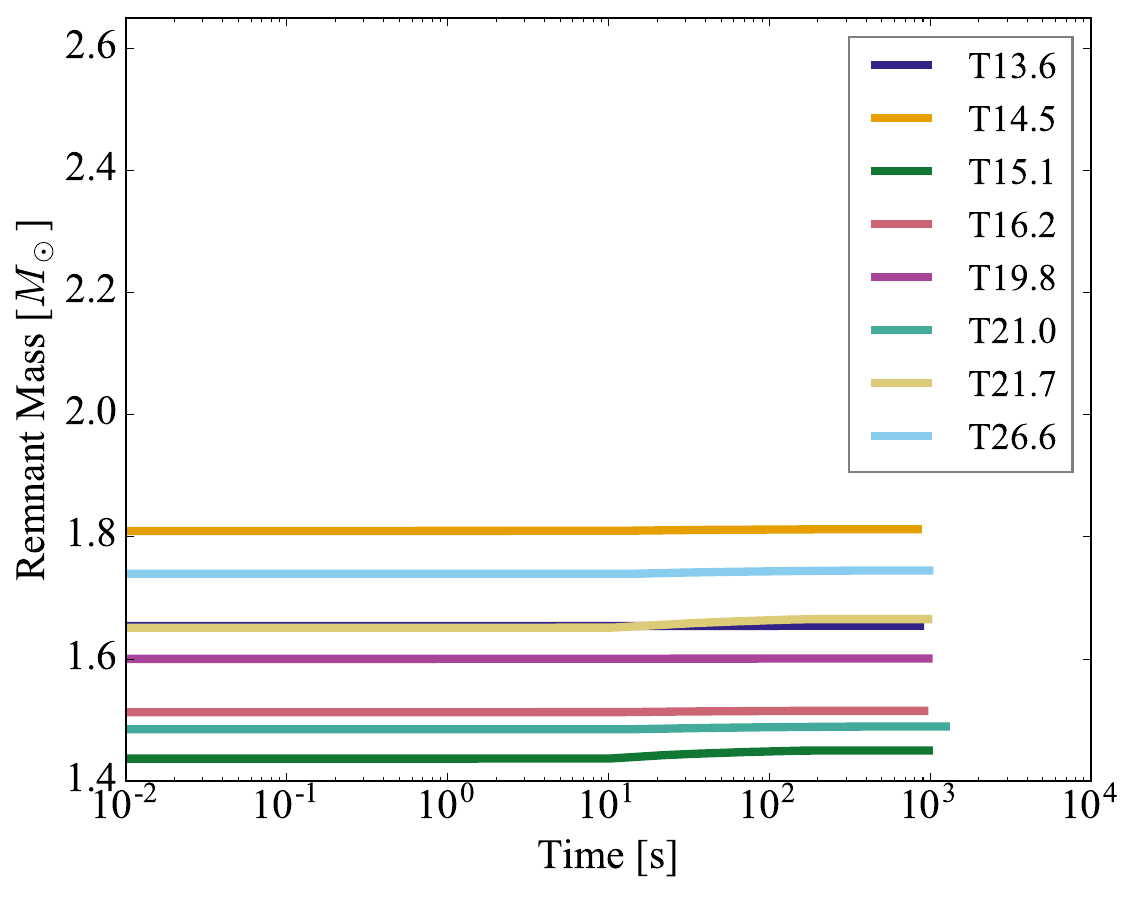}
 \end{center}
 \vspace{-12pt}
    \caption{Compact remnant masses as functions of post-bounce time ($t_\mathrm{pb}$) for all of our CCSN runs with different explosion mechanisms: neutrino-driven (top panel), classic piston-driven (middle panel), and thermal-bomb explosions with an energy deposition time of $1.0$\,s (bottom panel). Note that in the neutrino-driven explosions, all matter at densities above $10^{11}$\,g\,cm$^{-3}$ is considered as baryonic mass of the forming PNS, whereas in the other two cases the lines represent the mass coordinates of the inner boundaries of the computational grids. After the inner boundaries are opened, the lines in all simulations represent the baryonic mass that ends up in this central volume including the matter that has fallen through the open boundaries.}
    \label{fig:fallback}
\end{figure}

In Figure~\ref{fig:shells} this can be seen by the mass shells that initially fall inward through the shock, settle in a dense layer close to the inner boundary of the computational domain (yellow line), whose retraction they follow for a certain period of time, and then move outward with high velocity to trace the outward going shock. Figure~\ref{fig:fallback} displays this evolution in terms of the PNS mass as a function of time. In the top panel of this figure we see that the compact remnant's mass increases due to mass accretion for $\sim$1\,s, signalling the phase of shock stagnation. After the shock has started to move outward, the remnant mass starts to decrease slightly again, since some of its matter manages to become unbound in the neutrino-driven wind.

\begin{figure}
\begin{center}
	\includegraphics[width=0.98\columnwidth]{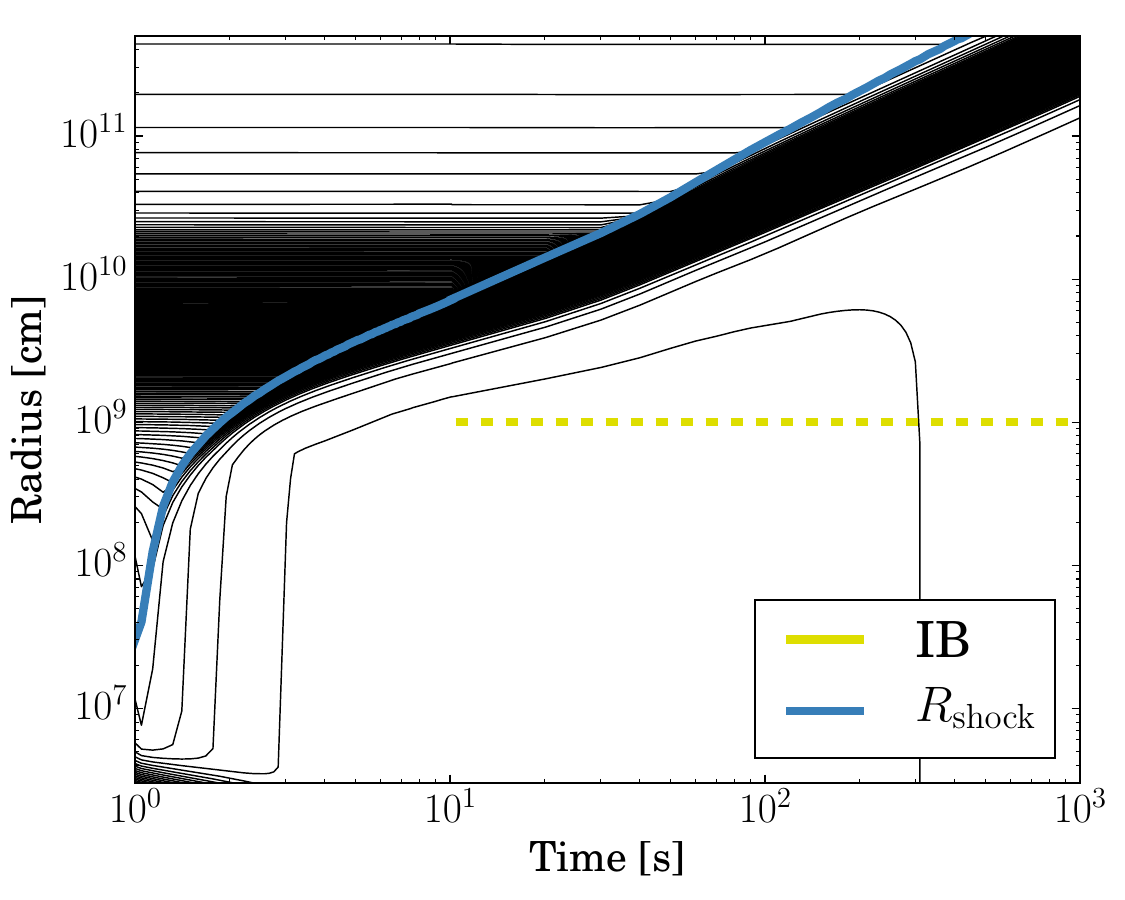}
 \vspace{-5pt}
	\includegraphics[width=0.98\columnwidth]{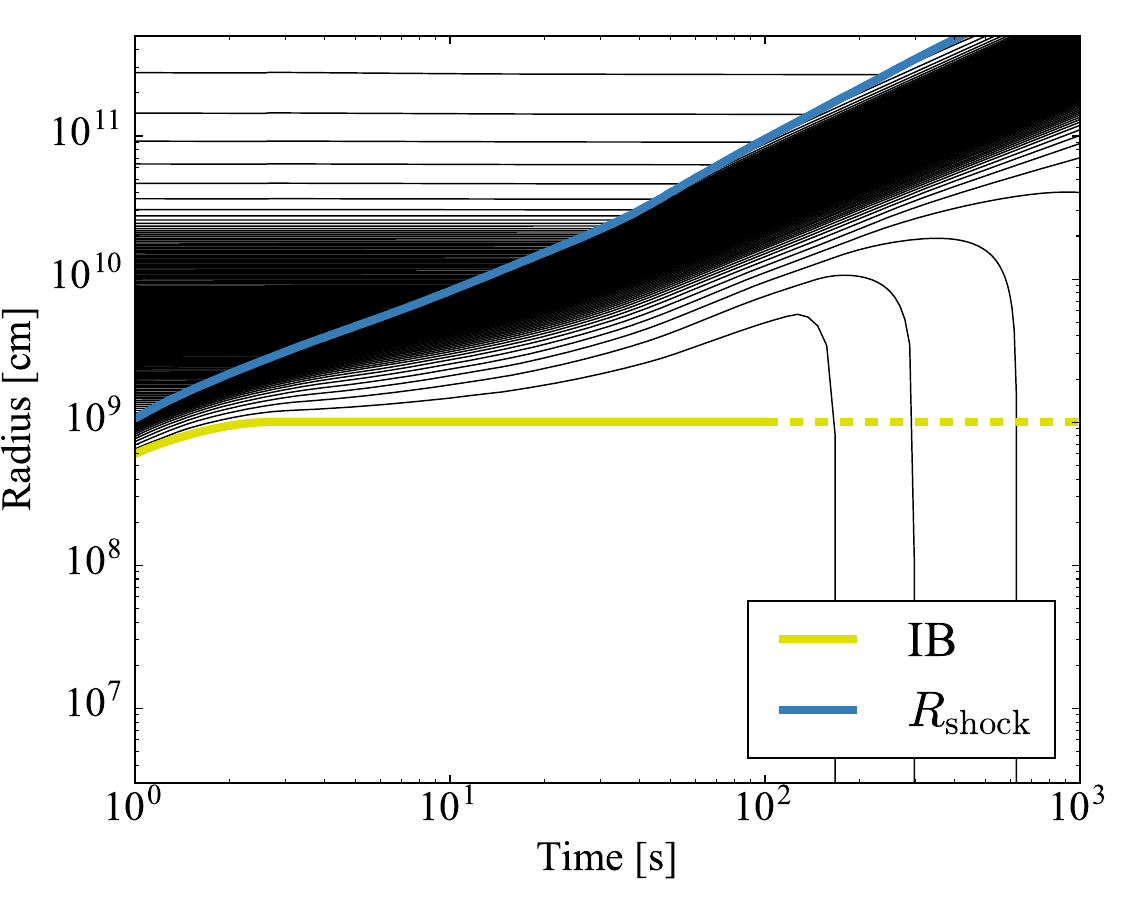}
 \vspace{-5pt}
	\includegraphics[width=0.98\columnwidth]{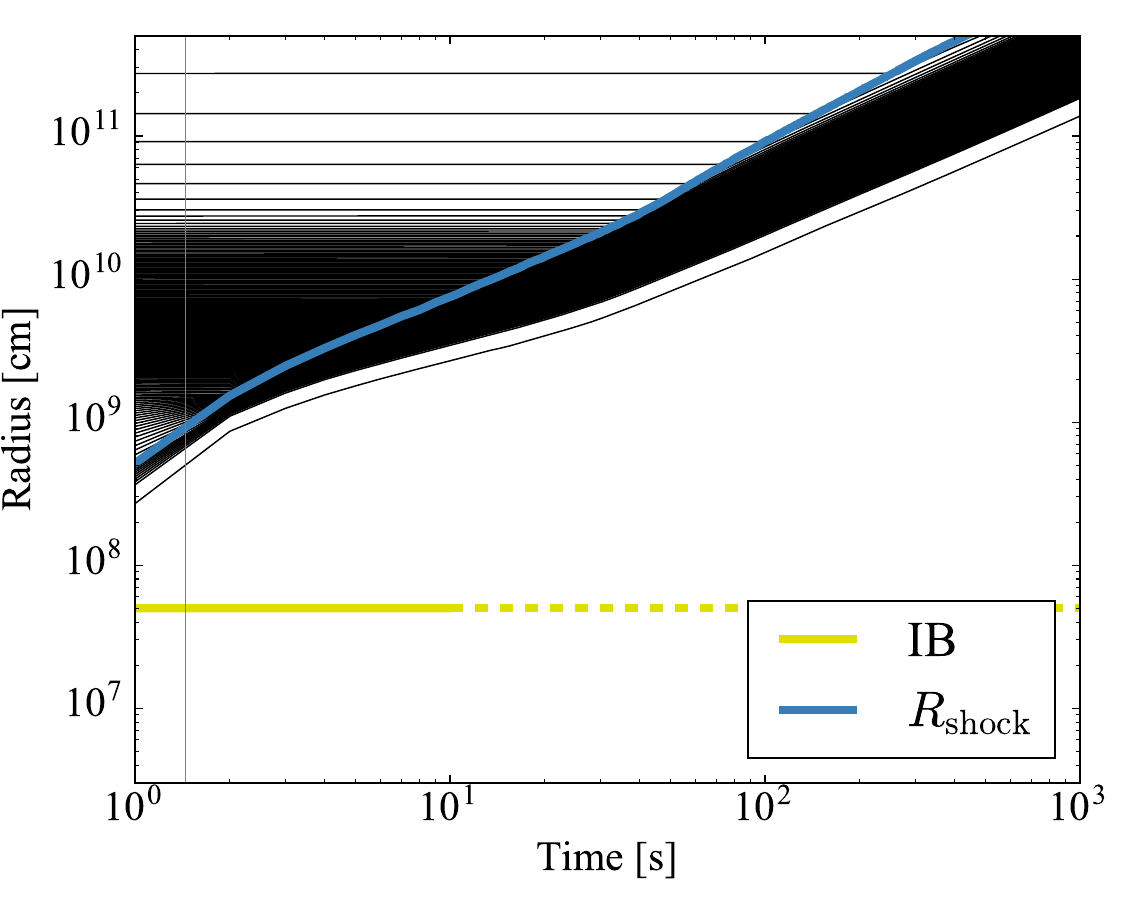}
 \end{center}
 \vspace{-5pt}
    \caption{Long-term radius evolution of Lagrangian mass shells with time (measured from the start of the simulations) for the 21.0\,$M_\odot$ progenitor with different explosion mechanisms (continuation of the evolution shown in Figure~\ref{fig:shells}): neutrino-driven (top panel), classic piston-driven (middle panel), and thermal bomb with the vertical line corresponding to the end of the energy deposition (bottom panel). The thin black solid lines are the mass shells, spaced in steps of 0.025\,$M_\odot$; the blue line marks the shock radius; the solid yellow line marks the closed inner grid boundary, and its end point the moment the boundary is opened, from where on the yellow line is dashed.}
    \label{fig:shells_xl}
\end{figure}

The situation is different in the piston-driven explosion (Figure~\ref{fig:shells}, middle panel). The core collapses until it reaches $500$\,km at 0.45\,s, when the shock forms. The movement of the shock radius (blue line) shows that there is no shock stagnation in this case, but that the shock is accelerated outward as soon as it has formed, being pushed by the outward motion of the inner grid boundary (yellow line). Between inner boundary and shock a significant amount of matter is driven outward while still staying close to the inner boundary. This leads to substantial later fallback of mass when the inner boundary is opened, which for instance affects more than $0.12\,M_\odot$ in the $21.0\,M_\odot$ explosion. Fallback will be discussed in more detail in the next subsection. Both the velocity of the shock and the matter density in the postshock region are higher than for the neutrino-driven explosion.

The collapse phase is simulated in exactly the same manner in the thermal-bomb model (bottom panel of Figure~\ref{fig:shells}) as in the piston-driven case (middle panel). The energy deposition sets in at the bounce at 0.45\,s and continues for 1.0\,s. The injection phase is indicated by the thin vertical lines. Since the energy is injected continuously, the shock accelerates less rapidly than in the piston-driven explosion, and the mass shells following the shock move outward more slowly, too, but still faster than in the neutrino-driven explosion. The inner boundary is held constant at $500$\,km with almost no matter accumulated in its close vicinity.

\begin{figure}
	\includegraphics[width=\columnwidth]{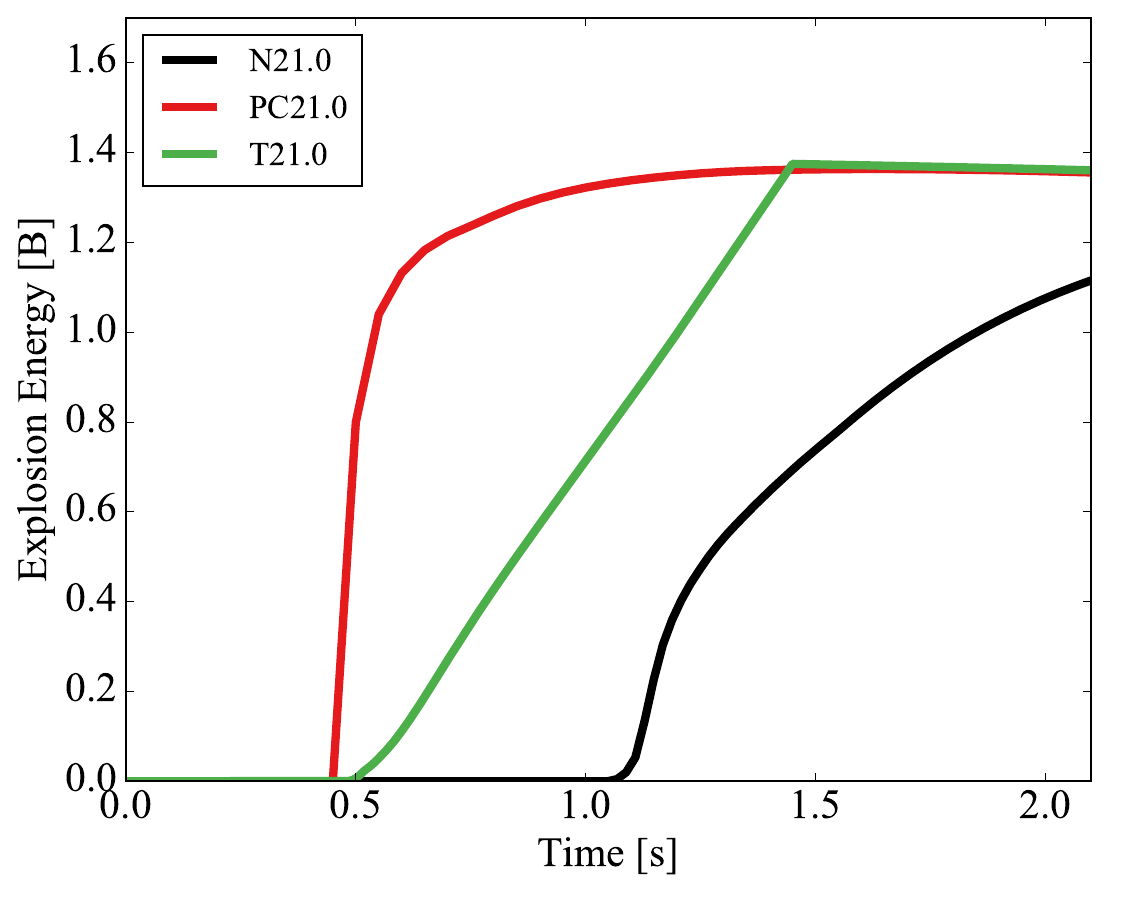}
	\includegraphics[width=\columnwidth]{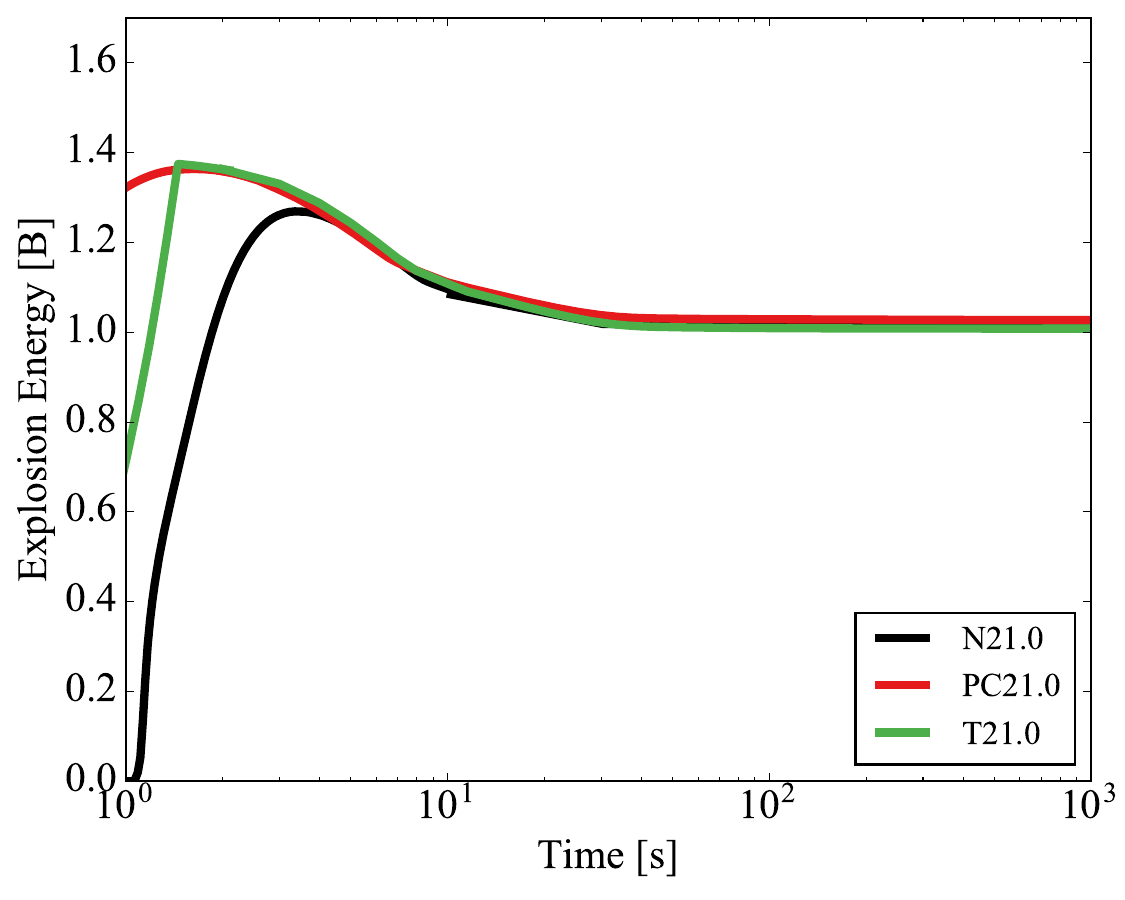}
 \vspace{-15pt}
	\caption{Early (top panel) and late (bottom panel) evolution of the explosion energy for the 21.0\,$M_\odot$ progenitor exploded with different mechanisms: neutrino-driven (black), classic piston-driven (red), and thermal-bomb explosion (green). Time is measured from the start of the simulations.}
    \label{fig:eexp}
\end{figure}

The shock radius at 2\,s is the largest in the piston-driven explosion, followed by the thermal bomb, and then by the neutrino-driven explosion. The reason is that the energy deposition is fastest by the piston, whereas it is slower by the thermal bomb and even slower by the neutrino heating with the neutrino engine. This is visible in Figure~\ref{fig:eexp}, which illustrates the early evolution of the explosion energy in the upper panel.

As we have the same explosion energy at infinity by design (visible in the lower panel of Figure~\ref{fig:eexp}), it can be expected that the shock will reach a similar radius in the end (see next subsection), accelerating just more slowly in cases with longer injection times. Moreover, the shock is revived later in the neutrino-driven explosion --\,at about 1\,s after the start of the collapse\,--, whereas it starts its rapid outward expansion already at 0.45\,s in the thermal-bomb and piston-driven explosions. This is a second reason why the shock radius is smaller at 2\,s in the neutrino-driven explosion.

During the collapse phase the inner grid boundary in both the thermal-bomb and the piston-driven explosions contracts to 500\,km, which is not entirely consistent with the behaviour witnessed for a corresponding mass shell in the neutrino-driven explosion. Note that the collapse phase in the P-HOTB models is computed for the entire iron core down to the center of the progenitor, and that the inner grid boundary, introduced only shortly after core bounce, is located at a mass shell that ends up deep inside the new-born PNS (see yellow line in the upper panel of Figure~\ref{fig:shells}). Therefore the inner boundaries of the thermal-bomb and piston models should better be compared to the red line in the upper panel of Figure~\ref{fig:shells}, which tracks the trajectory of the first mass shell that follows the expansion of the outward moving shock in the neutrino-driven explosion. This mass shell contracts to a deeper location at a smaller radius of only $\sim$150\,km. The choice of a 500\,km boundary in our thermal-bomb and piston simulations was influenced by the conventional recipes to facilitate comparisons with existing literature. In our previous study \citep{2023MNRAS.518.1818I} we also explored the option of contracting the inner boundary to 150\,km for thermal-bomb explosions, revealing no substantial variations in the results. However, the deeper contraction increased the computational demands of the simulations considerably. The influence of a deeper launching point of the outward going CCSN shock has more significant consequences in piston-driven explosions, where the discrepancies in the final explosion properties compared to thermal-bomb and neutrino-driven explosions can be important, as will be seen later. 

\subsubsection{Fallback phase}
\label{sec:fallphase}

The final explosion properties are determined only on longer evolution times of many seconds. Therefore we followed the long-term evolution in our 1D explosion models and will show that the artificial explosion mechanisms, in particular the piston mechanism, can result in late fallback of significant amounts of matter. This affects the masses of the compact remnants and the (observable) products of CCSN nucleosynthesis and is an unphysical effect, because it is not compatible with the behavior seen in neutrino-driven explosions. In this section we discuss the impact of fallback on the masses of the remnants, while its impact on the ejected $^{56}$Ni masses will be discussed in the next section.

\begin{figure}
	\includegraphics[width=\columnwidth]{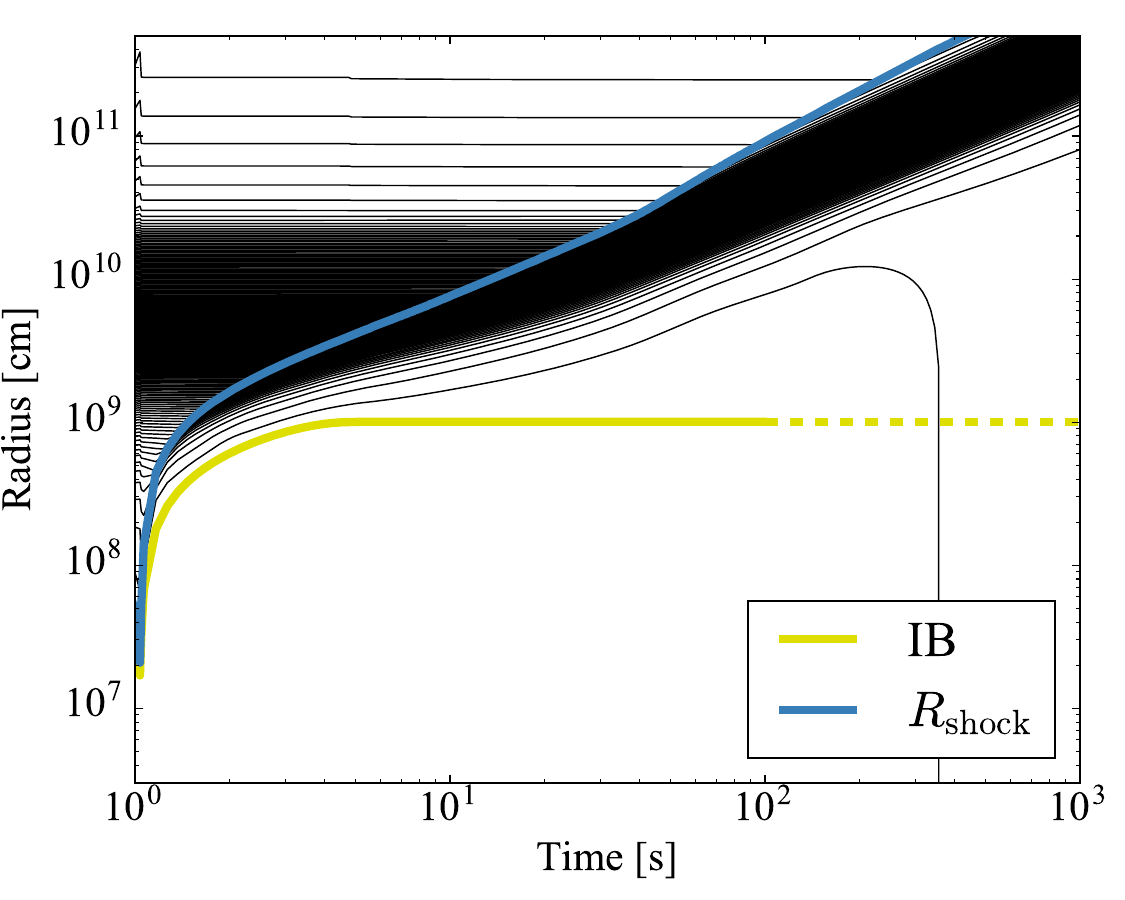}
  \vspace{-15pt}
\caption{Radius evolution of Lagrangian mass shells with time (measured from the start of the simulations) for the special-trajectory piston-driven explosion of the 21.0\,$M_\odot$ progenitor. The thin black solid lines are the mass shells, spaced in steps of 0.025\,$M_\odot$, and the blue line marks the shock radius, and the yellow line represent the inner grid boundary (i.e., the location of the piston), which is initially closed (solid line) and later open (dashed line).}
    \label{fig:special_trajectory_xl}
\end{figure}

\begin{figure}
	\includegraphics[width=\columnwidth]{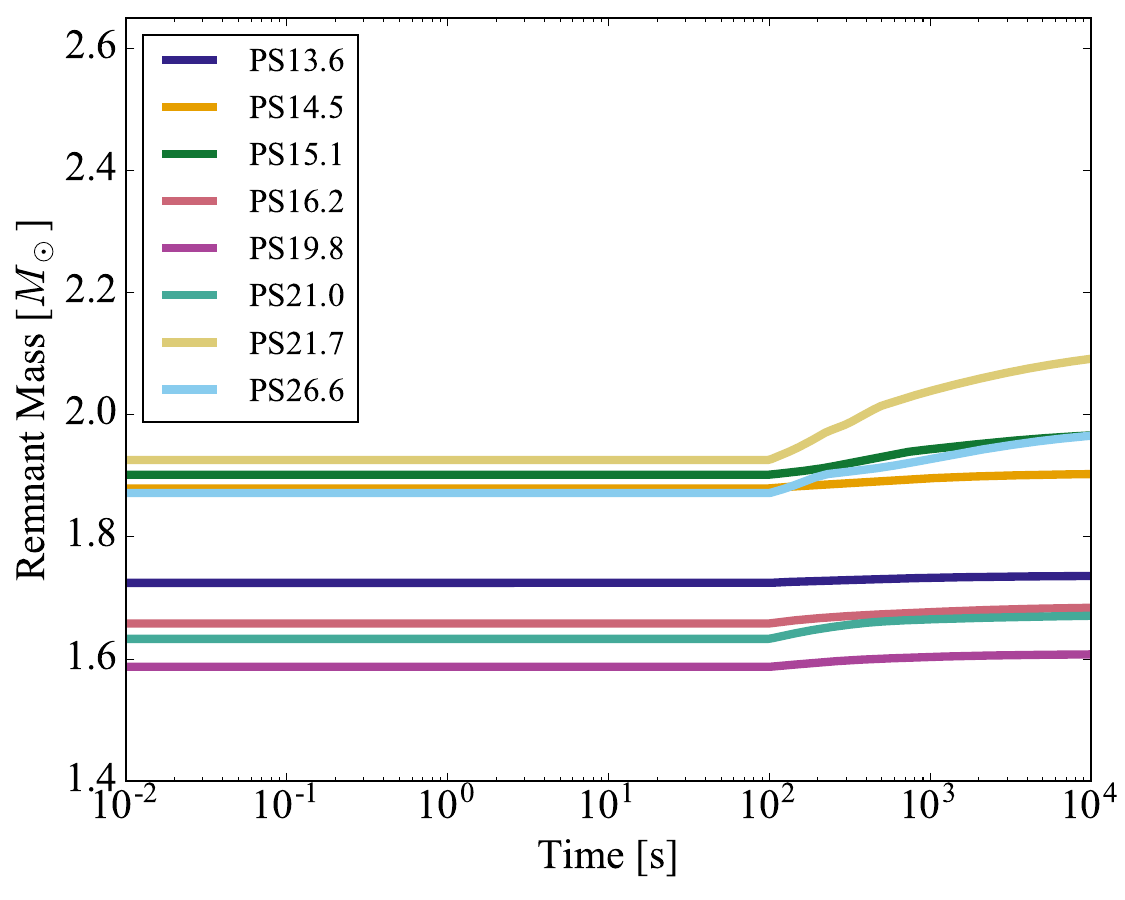}
  \vspace{-15pt}
\caption{Compact remnant masses as functions of post-bounce time for our CCSN runs of all progenitors with special-trajectory piston-driven explosions (analogous to Figure~\ref{fig:fallback}).}
    \label{fig:fallback_special}
\end{figure}

\begin{table*}
	\centering
	\caption{Results of the hydrodynamic simulations for the neutrino-driven explosions according to the P-HOTB method. The first column represents the mechanism $m$ and the ZAMS mass $M_{\rm ZAMS}$, the second column is the calibration model with the parameters from Table~\ref{tab:neutrino}, the third column is the final explosion energy $E_{\rm exp}$ (1\,B = 1\,bethe = $10^{51}$\,erg), the fourth column is the PNS mass $M_{\rm PNS}$ at 10\,s, the fifth column is the PNS mass $M_{\rm PNS}$ at $10^4$\,s, the sixth column is the amount of matter that falls back onto the PNS, $M_{\rm fb}$, the seventh and the eighth columns represent the final masses of $^{56}$Ni and $^{56}$Ni$+$tracer ejected in the explosions, $^{56}$Ni, and $^{56}$Ni+tracer, respectively.}
	\label{tab:neutrino_res}
  \begin{tabular}{ c|c|c|c|c|c|c|c }
  \hline
  \hline
  $mM_{\rm ZAMS}$ [$M_\odot$] & Calibration & $E_{\rm exp}$ [B] & initial $M_{\rm PNS}$ [$M_\odot$] & final $M_{\rm PNS}$ [$M_\odot$] &$M_{\rm fb}$ [$M_\odot$] & $M_{\rm Ni}$ [$M_\odot$] & $M_{\rm Ni+tr}$ [$M_\odot$] \\
\hline
N13.6  &   S19.8 &   1.88481 &   1.59318   & 1.59388  &  0.00069   &   0.05468 &   0.10983  \\
N14.5  &   W20   &   1.03172 &   1.80502   & 1.80686  &  0.00183   &   0.05356 &   0.06935  \\
N15.1  &   S19.8 &   0.55513 &   1.86263   & 1.86729  &  0.00466   &   0.04483 &   0.04926  \\
N16.2  &   W18   &   1.13908 &   1.57532   & 1.57783  &  0.00251   &   0.06015 &   0.09362  \\
N19.8  &   S19.8 &   2.04402 &   1.43082   & 1.43246  &  0.00164   &   0.05370 &   0.13217  \\
N21.0  &   W18   &   1.02704 &   1.54523   & 1.55031  &  0.00508   &   0.04819 &   0.08343  \\
N21.7  &   W18   &   0.63036 &   1.87384   & 1.89186  &  0.01802   &   0.06275 &   0.07578  \\
N26.6  &   W18   &   1.07301 &   1.78432   & 1.79888  &  0.01456   &   0.09541 &   0.13121  \\
\hline 
\end{tabular} 
\end{table*}

\begin{table*}
	\centering
	\caption{Results of the hydrodynamic simulations for the classic piston-driven explosions. Column headings have the same meaning as in Table~\ref{tab:neutrino_res}; additionally, $E_{\rm bind}$ (third column; 1\,B = 1\,bethe = $10^{51}$\,erg) is the gravitational binding energy of the matter on the computational grid in the pre-collapse progenitor. The initial mass of the PNS is the mass coordinate of the inner grid boundary (see the positions of the vertical dashed lines in Figure~\ref{fig:psn_closer}).}
	\label{tab:piston_res}
  \begin{tabular}{ c|c|c|c|c|c|c }
  \hline
  \hline
    $mM_{\rm ZAMS}$ [$M_\odot$] & $E_{\rm exp}$ [B] & $E_{\rm bind}$ [B] & initial $M_{\rm PNS}$ [$M_\odot$] & final $M_{\rm PNS}$ [$M_\odot$] &$M_{\rm fb}$ [$M_\odot$] & $M_{\rm Ni}$ [$M_\odot$] \\
 \hline
PC13.6  &   1.87626 &  $-$0.27411 &  1.65399  &   1.66203 & 0.00804   &   0.12125  \\
PC14.5  &   1.03601 &  $-$0.28223 &  1.80985  &   1.81794 & 0.00809   &   0.04486  \\
PC15.1  &   0.55359 &  $-$0.44536 &  1.43694  &   1.89262 & 0.45568   &   0.00000  \\
PC16.2  &   1.14409 &  $-$0.35633 &  1.51282  &   1.55515 & 0.04233   &   0.10314  \\
PC19.8  &   2.03636 &  $-$0.34489 &  1.60033  &   1.60806 & 0.00773   &   0.08136  \\
PC21.0  &   1.02714 &  $-$0.44649 &  1.48435  &   1.60841 & 0.12406   &   0.02771  \\
PC21.7  &   0.62949 &  $-$0.53520 &  1.65120  &   2.60606 & 0.95486   &   0.00000  \\
PC26.6  &   1.06576 &  $-$0.81031 &  1.73833  &   2.18437 & 0.44604   &   0.00000  \\
\hline 
\end{tabular} 
\end{table*}

\begin{table*}
	\centering
	\caption{Results of the hydrodynamic simulations for the thermal-bomb explosions. Column headings have the same meaning as in Table~\ref{tab:piston_res}.}
	\label{tab:thermal_res}
  \begin{tabular}{ c|c|c|c|c|c|c }
  \hline
  \hline
    $mM_{\rm ZAMS}$ [$M_\odot$] & $E_{\rm exp}$ [B] & $E_{\rm bind}$ [B] & initial $M_{\rm PNS}$ [$M_\odot$] & final $M_{\rm PNS}$ [$M_\odot$] &$M_{\rm fb}$ [$M_\odot$] & $M_{\rm Ni}$ [$M_\odot$] \\
 \hline
T13.6  &  1.84431 &  $-$0.27411 & 1.65399 & 1.65463 & 0.00064   &   0.08132  \\
T14.5  &  1.04399 &  $-$0.28223 & 1.80985 & 1.81261 & 0.00276   &   0.03122  \\
T15.1  &  0.56188 &  $-$0.44536 & 1.43694 & 1.45027 & 0.01333   &   0.20690  \\
T16.2  &  1.13944 &  $-$0.35633 & 1.51282 & 1.51521 & 0.00239   &   0.08945  \\
T19.8  &  2.05990 &  $-$0.34489 & 1.60033 & 1.60101 & 0.00069   &   0.03726  \\
T21.0  &  1.00807 &  $-$0.44649 & 1.48435 & 1.48959 & 0.00524   &   0.08531  \\
T21.7  &  0.63432 &  $-$0.53520 & 1.65120 & 1.66540 & 0.01420   &   0.09249  \\
T26.6  &  1.04394 &  $-$0.81031 & 1.73833 & 1.74519 & 0.00686   &   0.08806  \\
\hline 
\end{tabular} 
\end{table*}

\begin{table*}
	\centering
	\caption{Results of the hydrodynamic simulations for the special-trajectory piston-driven explosions. Column headings have the same meaning as in Table~\ref{tab:piston_res}. Note that compared to the classic piston-driven models of Table~\ref{tab:piston_res}, the initial mass cut, i.e., the initial PNS mass and thus location of the inner grid boundary, is different.}
	\label{tab:piston_special_res}
  \begin{tabular}{ c|c|c|c|c|c|c }
  \hline
  \hline
    $mM_{\rm ZAMS}$ [$M_\odot$] & $E_{\rm exp}$ [B]  & $E_{\rm bind}$ [B]& initial $M_{\rm PNS}$ [$M_\odot$] & final $M_{\rm PNS}$ [$M_\odot$] &$M_{\rm fb}$ [$M_\odot$] & $M_{\rm Ni}$ [$M_\odot$] \\
 \hline
PS13.6  & 1.88266 & $-$0.22559  & 1.72472  & 1.73619  & 0.01141  & 0.07876  \\
PS14.5  & 1.02361 & $-$0.25908  & 1.87946  & 1.90306  & 0.02360  & 0.06941  \\
PS15.1  & 0.55550 & $-$0.26891  & 1.90184  & 1.96658  & 0.06474  & 0.04917  \\
PS16.2  & 1.13178 & $-$0.32734  & 1.65821  & 1.68409  & 0.02588  & 0.07811  \\
PS19.8  & 2.03560 & $-$0.36082  & 1.58732  & 1.60769  & 0.02037  & 0.07948  \\
PS21.0  & 1.03099 & $-$0.40926  & 1.63314  & 1.67100  & 0.03786  & 0.05436  \\
PS21.7  & 0.62011 & $-$0.48994  & 1.92619  & 2.09218  & 0.16599  & 0.00023  \\
PS26.6  & 1.07460 & $-$0.81977  & 1.87189  & 1.96603  & 0.09414  & 0.08524  \\
\hline 
\end{tabular} 
\end{table*}

In Figure~\ref{fig:fallback} the long-term evolution of the PNS mass is shown for the three explosion mechanisms. The PNS mass is either defined as the mass of matter at densities $\rho > 10^{11}$\,g\,cm$^{-3}$ for the neutrino-driven explosions before 10\,s after core bounce, or by the mass enclosed by the inner boundary of the computational grid for the piston and thermal-bomb models and for the neutrino-driven explosions at times later than 10\,s after bounce.

In the neutrino-driven explosion models, the PNS mass initially increases during the formation of the compact remnant by the accretion of matter from the collapsing stellar core (Figure~\ref{fig:fallback}, top panel). When the explosion sets in, the outward expansion of the CCSN shock quenches this accretion. Subsequently, the neutrino-driven wind blows out near-surface material of the PNS and leads to a more or less significant decrease of the remnant's mass, before at late post-bounce times ($t_\mathrm{pb} > 10$\,s) the PNS mass can again slightly increase by late-time fallback of gas that does not become gravitationally unbound in the explosion. In the piston-driven and thermal-bomb explosions, the remnant mass is constant and equal to the mass coordinate of the inner grid boundary until fallback sets in (at $t_\mathrm{pb} > 100$\, for the piston models and $t_\mathrm{pb} > 10$\,s for the thermal-bomb models; Figure~\ref{fig:fallback}, middle and bottom panels, respectively).  

In the neutrino-driven explosion models, the inner grid boundary for the long-term evolution is moved to $10^9$\,cm at 10\,s post bounce and switched to an open (outflow) boundary (Figure~\ref{fig:shells_xl}, top panel, for the case of the 21.0\,$M_\odot$ explosion model). Later on some matter manages to fall back through the open boundary to be accreted onto the PNS, but the effect is rather insignificant, especially if compared to the substantial fallback in the piston-driven explosion, setting in immediately after the inner grid boundary is opened at 100\,s (Figures~\ref{fig:fallback} and~\ref{fig:shells_xl}, middle panels). For the thermal-bomb explosion, the inner boundary is opened at 10\,s (Figure~\ref{fig:shells_xl}, bottom panel). Afterwards there is some fallback, but the amount is rather small and in fact quite similar to the neutrino-driven case. Quantitative results for all of our simulations with the different explosion mechanisms are listed in Tables~\ref{tab:neutrino_res}--\ref{tab:thermal_res}.

The \textit{neutrino-driven explosions} have little fallback, because the continuous (though decaying with time) energy input by neutrinos provides a long-lasting outward accelerating force, which pushes matter to large radii. A similar effect is at work in the thermal-bomb models. 

The reason for the much larger fallback in \textit{piston-driven explosions} becomes evident by comparing the long-term evolution of the mass shells for the different explosion mechanisms; this is illustrated by Figure~\ref{fig:shells_xl}. The radius of the inner boundary in the piston-driven explosion is kept constant after it has reached $10^9$\,cm, which is rather far out in comparison to the thermal bomb explosions, where it is kept constant at $5\cdot 10^7$\,cm. The open boundary at the huge radius in piston-driven models makes the fallback so extreme, in particular since a lot of matter stays close to the grid boundary. Consequently, the fallback mass in piston-driven explosions is vastly overestimated compared to neutrino-driven explosions.

The amount of fallback depends both on the binding energy of the overlying material and on the explosion energy. The progenitors with masses $14.5$, $16.2$, $21.0$, and $26.6\,M_\odot$ have similar explosion energies of $\sim$1\,B, but their binding energies differ ($-0.28$, $-0.36$, $-0.45$, and $-0.81$~B, respectively); Tables~\ref{tab:piston_res}--\ref{tab:piston_special_res} list these values for the gravitational binding energies of all matter on the computational grid in the pre-collapse progenitors (thus excluding the progenitor cores that are assumed to become the initial compact remnants). A fraction of the deposited explosion energy goes into unbinding the overlying material. At roughly the same explosion energy, progenitors with higher binding energy typically experience more fallback.

The dependence on the explosion energy is evident from the following comparison: The progenitors of 15.1 and 21.7\,$M_\odot$ have explosion energies of $\sim$0.6\,B, whereas the progenitors with 13.6 and 19.8\,$M_\odot$ have explosion energies of $\sim$2\,B. The higher the explosion energy, the more energy is deposited in the model, resulting in less matter falling back onto the PNS. Therefore, with a lower explosion energy a given progenitor will experience more fallback and produce a PNS with higher mass, which explains the tendency that progenitors with lower explosion energies yield higher final PNS masses  (cf.\ Figure~\ref{fig:fallback} and  Table~\ref{tab:piston_res}).

The same trends, although much weaker, are present in the \textit{thermal-bomb models} (see the small changes of the remnant masses in Figure~\ref{fig:fallback}, lower panel). Quantitative information is provided by the fallback masses $M_{\rm fb}$ listed in Table~\ref{tab:thermal_res}. The fallback here is significantly less extreme than in the models with piston mechanism because of the differences in the treatment of the inner boundary and of the energy transfer to the explosions by thermal-energy deposition instead of the kinetic push of the outward moving piston. The thermal energy injection into a defined mass interval $\Delta M$ just outside the inner grid boundary drives the expansion of this mass shell and of the overlying layers, thus thinning out the region around the boundary at a fixed, much smaller radius than the grid boundary in the piston models. Since the binding energies as well as the locations of the initial mass cuts are the same in the piston and corresponding thermal-bomb models, and the explosion energies were calibrated to the same values, the differences of the progenitor structures and explosion energies cause the same overall tends in the fallback masses, though on a much lower level in the thermal-bomb explosions.
The amount of fallback in the thermal-bomb models could potentially also depend on the energy-deposition timescale $t_\mathrm{inj}$. While the models discussed here were computed with $t_\mathrm{inj}=1.0$\,s, we checked that replacing this by $t_\mathrm{inj}=0.2$\,s or $t_\mathrm{inj}=2.0$\,s does not lead to significant changes of the fallback masses for any of the thermal-bomb explosions considered in this work \citep[see also][]{2023MNRAS.518.1818I}.

The classic piston explosions thus massively overestimate the amount of fallback. This, however, can be avoided by using the special trajectories discussed in Section~\ref{sec:piston} instead of the classical trajectories. The results of our simulations for special-trajectory piston-driven explosions with the parameters listed in Table~\ref{tab:piston_special_par} are presented in Figures~\ref{fig:special_trajectory_xl} and~\ref{fig:fallback_special}, and the corresponding quantitative information is provided in Table~\ref{tab:piston_special_res}. Figure~\ref{fig:special_trajectory_xl} shows that the inner grid boundary, which represents the piston location, now contracts to a smaller minimum radius and then expands to the same coasting radius of $10^9$\,cm as before in the middle panel of Figure~\ref{fig:shells_xl}. Despite the fact that the inner boundary is again opened at 100\,s after bounce, comparing the mass shells in both figures already suggests considerably less fallback in the special-trajectory explosion of the 21\,$M_\odot$ progenitor. This conclusion is confirmed by Figure~\ref{fig:fallback_special}, where the compact remnant masses of all special-trajectory piston models display far less increase at post-bounce times $t_\mathrm{pb} > 100$\,s than the corresponding classic piston-driven explosions in the middle panel of Figure~\ref{fig:fallback}. Obviously, the use of the special trajectories results in a more reasonable amount of fallback, more similar to that in the neutrino-driven explosions, mainly because less matter stays close to the inner boundary at late times. 

This dynamical difference is a consequence of several aspects in the special-trajectory piston setup. Besides the initial contraction of the inner grid boundary to a deeper radius, the infall phase is also longer, because the special trajectory is located at a significantly larger mass coordinate than the original classical trajectory (compare the initial PNS masses for the 21.0\,$M_\odot$ models in Tables~\ref{tab:piston_res} and~\ref{tab:piston_special_res}). This difference in the piston position, which is motivated and guided by the neutrino-driven explosion model, has the consequence that the special-trajectory piston pushes into lower-density matter and exerts its push over a larger distance, starting from a point deeper in the gravitational potential of the PNS. All these factors reduce the mass in the vicinity of the inner boundary when the latter is opened at 100\,s after bounce. Note that the choice of the special trajectories by the first mass shells following the onset of outward shock expansion in the neutrino-driven models implies that the mass coordinate of these trajectories also defines the PNS mass in the P-HOTB models at that time. Despite this connection, however, the initial PNS masses in the special-trajectory piston-driven explosions are generally higher than the initial PNS masses for the P-HOTB explosions listed in Table~\ref{tab:neutrino_res}. This difference can be understood by the fact that the PNS masses in Table~\ref{tab:neutrino_res} are measured at 10\,s and the PNSs have lost some mass in the neutrino-driven wind that sets in after shock revival (see discussion for the upper panel of Figure~\ref{fig:fallback}).

\begin{figure}
	\includegraphics[width=\columnwidth]{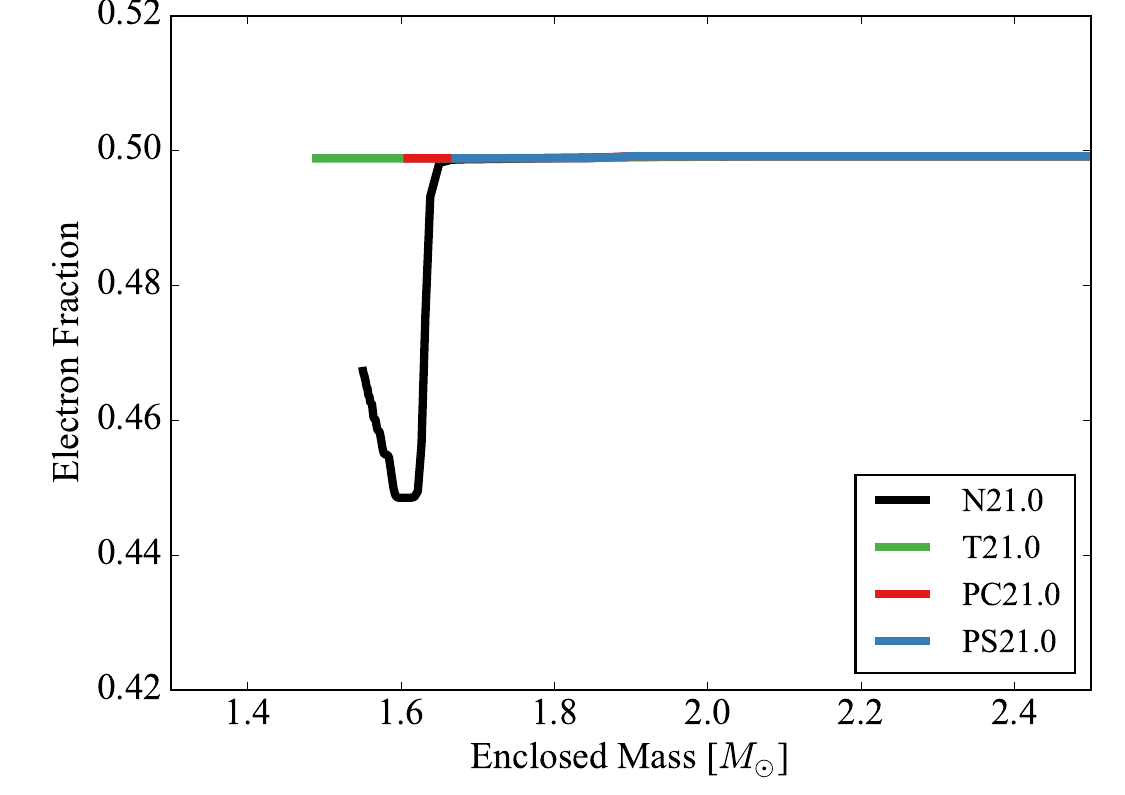}
    \vspace{-15pt}
    \caption{Electron fractions $Y_e$ versus enclosed mass at the end of our CCSN simulations (1400\,s for the thermal-bomb explosion (T21.0) and $10^4$\,s for the neutrino-driven (N21.0), classic piston-driven (PC21.0), and special-trajectory piston-driven (PS21.0) explosions) for the 21.0\,$M_\odot$ progenitor, blown up with all considered explosion mechanisms.}
    \label{fig:ye}
\end{figure}

\begin{figure*}
	\includegraphics[width=\columnwidth]{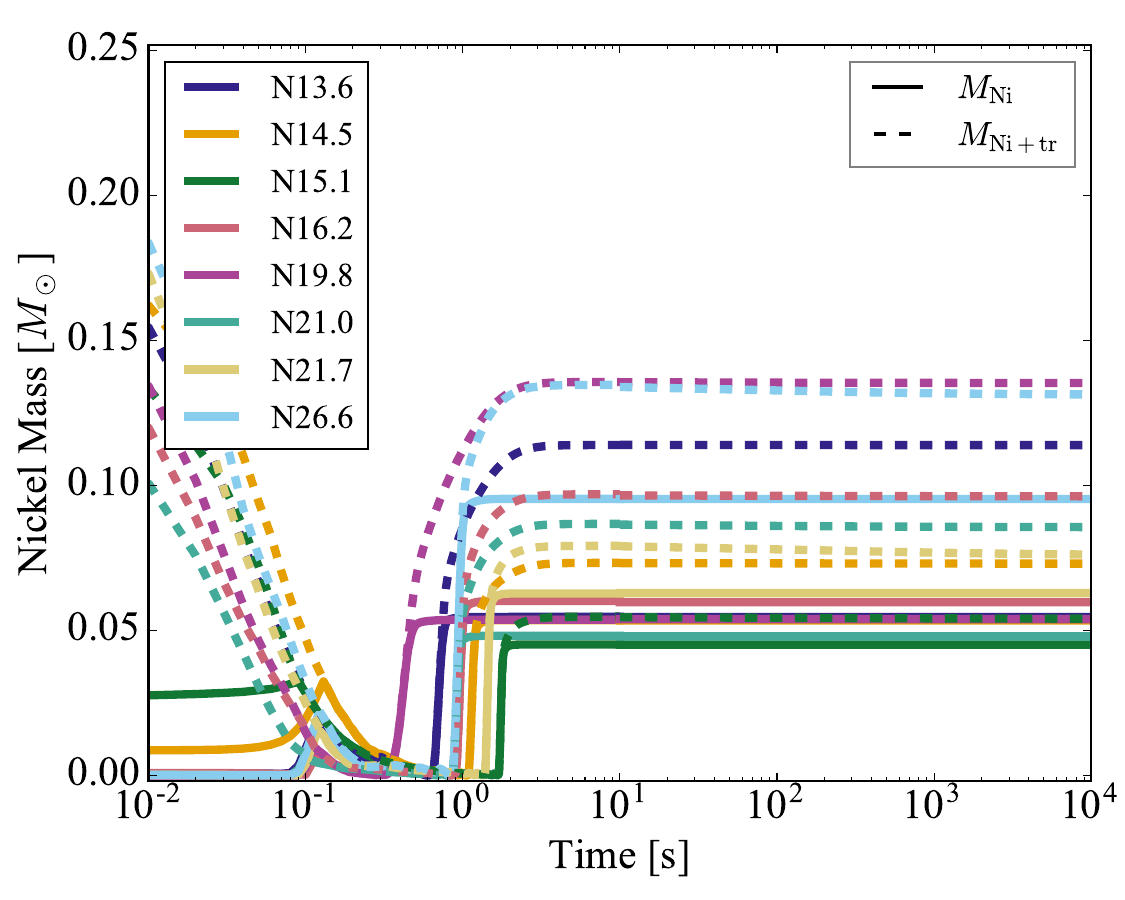}
 	\includegraphics[width=\columnwidth]{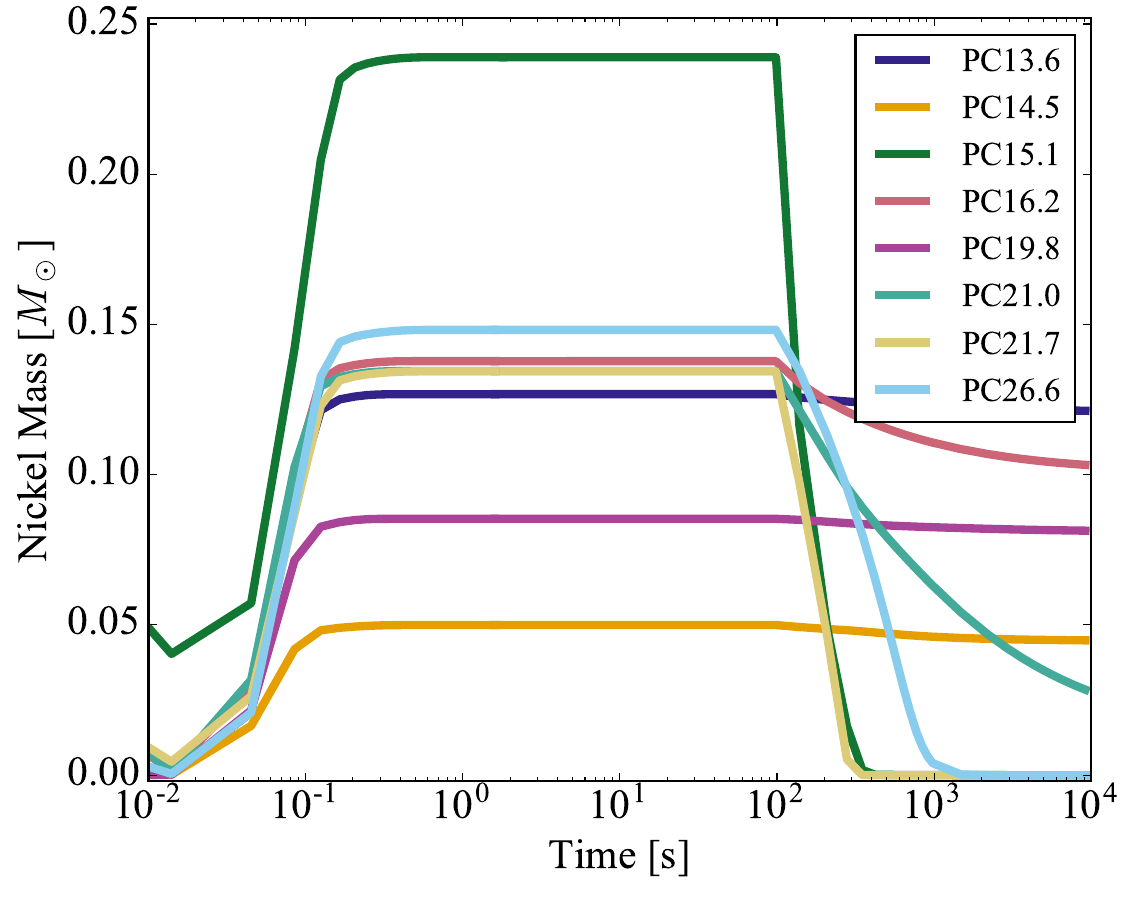}
 	\includegraphics[width=\columnwidth]{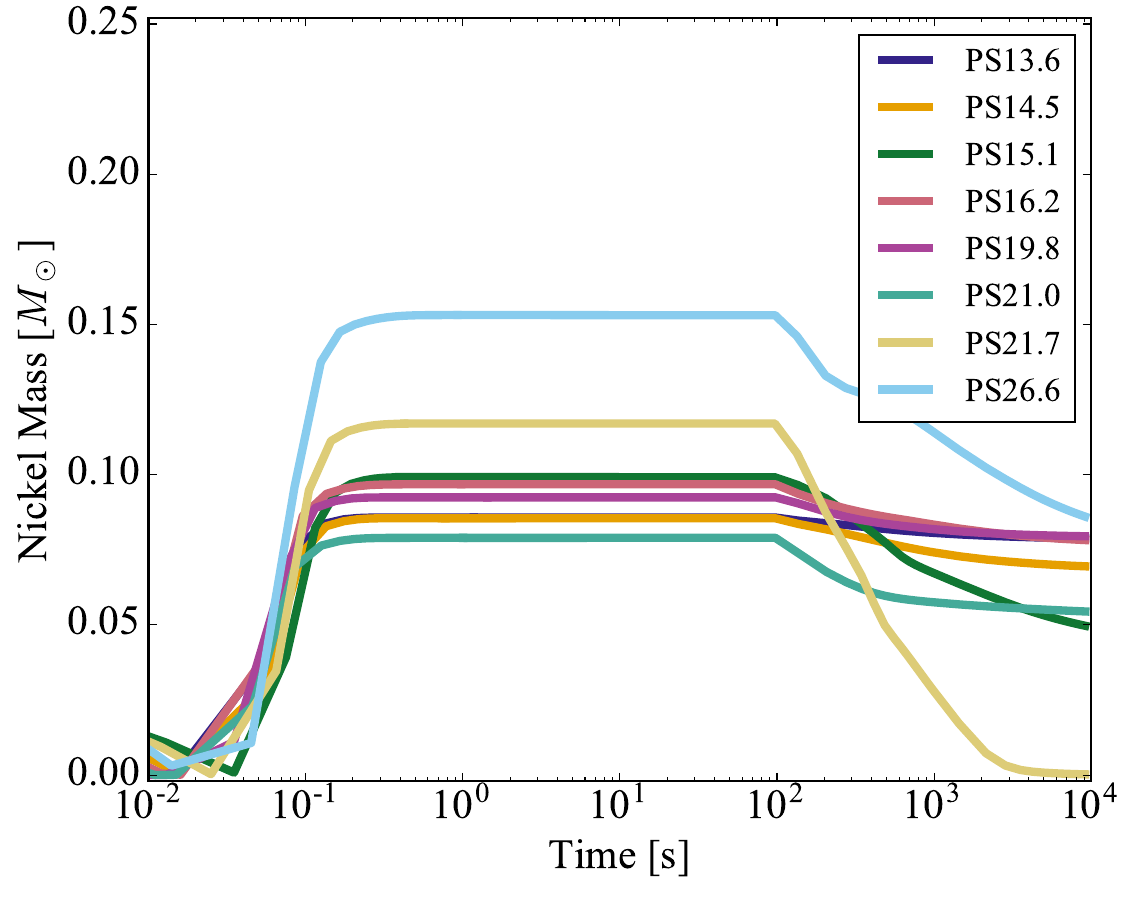}
	\includegraphics[width=\columnwidth]{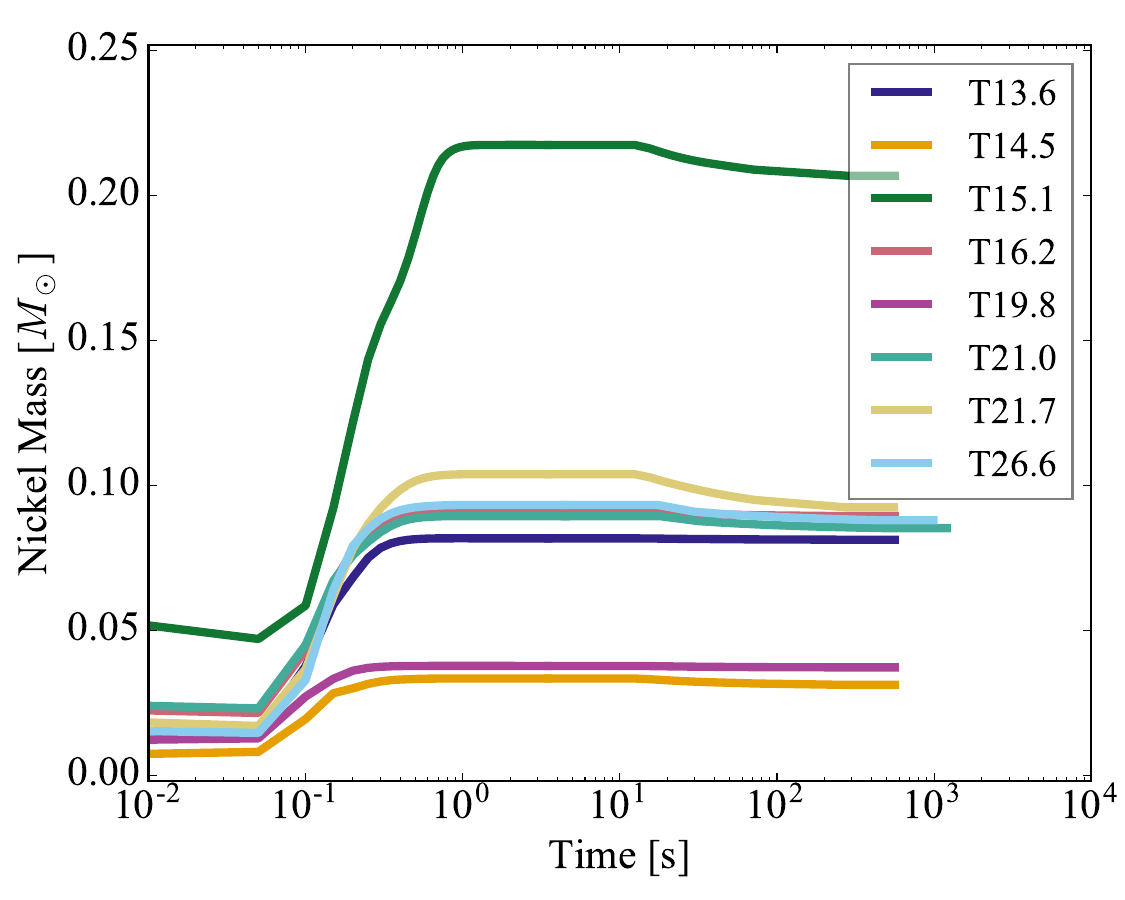}
    \caption{$^{56}$Ni masses as functions of post-bounce time for all of our CCSN models with different explosion mechanisms: neutrino-driven (top left panel, also showing the mass of the neutron-rich tracer nucleus), classic piston-driven (top right panel), special-trajectory piston-driven (bottom left panel), and thermal-bomb explosions with an energy-deposition time of 1.0\,s (bottom right panel). Note that in the neutrino-driven models the lines show $^{56}$Ni and tracer masses in ejected matter only after the onset of the explosions between about 0.4\,s and 2\,s after the start of the simulations, corresponding to the steep rise of the lines after the minima. In the piston-driven models the inner boundary is opened and fallback sets in at 100\,s after bounce, whereas both happens in the neutrino-driven and thermal-bomb explosions at 10\,s post bounce. Fallback is very small in the neutrino-driven explosions (see also Figure~\ref{fig:summary_neutrino}).}
    \label{fig:ni}
\end{figure*}

\subsection{Nickel production}
\label{sec:niprod}

We restrict the discussion of explosive nucleosynthesis in our 1D CCSN models to the production of $^{56}$Ni as a representative iron-peak nucleus. The nickel masses as functions of time after bounce (post-bounce time $t_\mathrm{pb}$) are presented in Figure~\ref{fig:ni}. The upper left panel corresponds to the neutrino-driven mechanism. Solid lines show the $^{56}$Ni masses for the explosion runs of all progenitors, and dashed lines the summed masses of $^{56}$Ni and of our tracer nucleus representing the production of neutron-rich species in the neutrino-driven explosions (see Section~\ref{sec:networks}). The final nickel mass consists of nickel produced in shock-heated and in neutrino-heated matter. The tracer is only made in matter that has experienced $\nu_e$ and $\bar\nu_e$ reactions (emission and/or absorption) and thereby has become neutron-rich. Such conditions are absent in piston-driven and thermal-bomb explosions, where neutrino reactions are not included. Therefore $Y_e$ in the ejecta has the values (close to 0.5) of the progenitors in the mass shells outside the initial mass cut (see Figure~\ref{fig:ye}), and the tracer mass is zero in those models.

The tracer material mainly forms when neutrino-heated, neutron-rich ejecta that are initially at NSE temperatures experience expansion cooling and thus nucleon recombination to $\alpha$-particles and heavy nuclei during the freeze-out or alpha-rich freeze-out from NSE. Therefore the production of the tracer depends on the electron fraction in this material, which, however, is uncertain in our P-HOTB models because of the approximate treatment of the neutrino transport and a highly sensitive dependence of $Y_e$ on the radiated neutrino luminosities and spectra. For this reason some of the tracer material could be $^{56}$Ni instead, which motivates us to consider the mass of $^{56}$Ni as a lower limit to the production of this isotope and the summed yields of $^{56}$Ni and tracer, $M_{\rm Ni+tr}$, as an upper limit for the ejected $^{56}$Ni mass \citep[see the discussion by][]{2016ApJ...821...38S}. We note in passing, however, that 1D neutrino-driven explosions are not able to adequately capture the nucleosynthesis conditions in the innermost ejecta of 3D models of neutrino-driven explosions \citep[e.g.,][]{Wongwathanarat+2013,Wongwathanarat+2017,2016ApJ...818..123B,Wanajo+2018,Sieverding+2023,Wang+2024a,Wang+2024b}. Because of differences in the electron fraction, entropy, and expansion history, precise predictions of iron-group and trans-iron abundances cannot be expected from the 1D simulations. For example, mass elements ejected at a certain time in 1D models possess a single value of $Y_e$, entropy, and expansion time scale, whereas in 3D the ejected matter has a distribution of different conditions at each moment. Moreover, even some of the matter surviving as $\alpha$-particles in 1D during the freeze-out from NSE might instead form $^{56}$Ni in 3D explosions, because the entropies in neutrino-heated ejecta are overestimated in 1D models \citep[we refer to the corresponding discussion by][]{2020ApJ...890...51E}. Therefore, at best, the overall magnitude of the yields and their general trends (e.g., variations with progenitor and explosion energy) estimated on grounds of 1D neutrino-driven CCSN simulations might be similar to the multi-dimensional results.

We find that the nickel masses in our P-HOTP simulations for all progenitors lie between 0.045\,$M_\odot$ (minimal $^{56}$Ni yield) and 0.132\,$M_\odot$ (maximum value of $M_{\rm Ni+tr}$; see Figure~\ref{fig:ni}, upper left panel, and Table~\ref{tab:neutrino_res}). These values are indeed compatible with the $^{56}$Ni production obtained in similarly energetic, self-consistent 3D simulations \citep{Sieverding+2023,Wang+2024a,Wang+2024b}. Note that the $^{56}$Ni and tracer masses seen in the upper left panel of Figure~\ref{fig:ni} before the onset of the explosions between $\sim$0.4\,s and $\sim$2\,s after bounce (i.e., before the steep increase after the minimum of the curves) are irrelevant when discussing ejecta. This pre-explosion nickel and tracer material is in the outer layers of the iron core of the progenitor when they are still infalling before and early after core bounce. It will reach NSE once it has passed the CCSN shock. In NSE, the nuclei will be dissociated into nucleons and $\alpha$-particles. Some of this matter will get accreted onto the PNS, some of it will be re-ejected and then recombine to heavy nuclei in a nuclear freeze-out.

The upper right panel of Figure~\ref{fig:ni} contains the $^{56}$Ni masses for the classic piston-driven explosions. All nickel is produced in the first 0.1--0.3\,s after bounce, and the yields are more or less comparable with nickel plus tracer in the neutrino-driven explosions. The 15.1\,$M_\odot$ explosion is an exception; it overproduces nickel because of the small PNS mass and thus the mass cut being deep inside the progenitor at high densities (see Figure~\ref{fig:psn_closer}, upper right panel). However, the picture changes after 100\,s, when the inner boundary is open and fallback takes place. In the explosions of the 15.1, 21.7, and 26.6\,$M_\odot$ stars, all nickel produced falls back onto the PNS. These cases have the highest fallback masses because of their low explosion energies (15.1 and 21.7\,$M_\odot$ progenitors) or because of the high binding energy (26.6\,M$_\odot$ progenitor). The amount of nickel falling back is clearly overestimated and does not agree with the neutrino-driven explosions. The piston-driven explosions of the 16.2 and 21.0\,$M_\odot$ stars have less fallback, although the ejected $^{56}$Ni mass still gets noticeably reduced. In the 13.6, 14.5, 19.8\,$M_\odot$ explosions the ejected nickel masses are not affected that much by fallback because of the high explosion energies and low binding energies, therefore the $^{56}$Ni masses are still in an acceptable range, when comparing them to the neutrino-driven explosions. 

Using instead the special-trajectory piston can improve the situation (Figure~\ref{fig:ni}, lower left panel). The amount of fallback is more reasonable than for the classic-piston cases. Only in the 21.7\,$M_\odot$ explosion again almost all of the initially produced nickel falls back onto the PNS (Table~\ref{tab:piston_special_res}), since this model simultaneously has a low explosion energy and a high binding energy. For the same reason we also still find a significant reduction of the $^{56}$Ni by fallback in the special-trajectory explosion of the 26.6\,$M_\odot$ progenitor, although the finally ejected nickel mass is still high.

In the thermal-bomb explosions (Figure~\ref{fig:ni}, lower right panel) almost all of the nickel is produced in the first 0.2--1.0\,s. This is longer than in the piston-driven explosions, since the energy in the thermal-bomb models is deposited more slowly over $t_\mathrm{inj}=1.0$\,s. The thermal bombs initially produce $^{56}$Ni masses (before fallback) in the same order of the progenitors as in the classic-piston-driven explosions. However, the yields are typically only about 40\%--80\% of the amounts made in the classic-piston models, and in the majority of cases closer to the nucleosynthesized nickel masses in the special-trajectory models. In none of the thermal-bomb explosions fallback plays a significant role in determining the ejected nickel masses. The 15.1\,$M_\odot$ explosion again produces too much nickel compared to the neutrino-driven explosion, because this model has the lowest PNS mass of all models. This means that the mass cut is located deep inside the progenitor, implying a high matter density just outside of the inner grid boundary (Figure~\ref{fig:psn_closer}, upper right panel), for which reason a lot of matter is shock-heated and produces $^{56}$Ni. In contrast, the 14.5 and 19.8\,$M_\odot$ explosions underproduce nickel compared to the neutrino-driven explosions. In the 14.5\,$M_\odot$ case this can be explained by the fact that this model has the largest PNS mass and therefore a very low matter density in the vicinity of the mass cut (see the location of the inner grid boundary in the upper right panel of Figure~\ref{fig:psn_closer}). Similarly, also the 19.8\,$M_\odot$ progenitor has a very low density close to the inner grid boundary because of its steep density drop at the $s/k_\mathrm{B} = 4$ location, which explains why its explosion produces a very low $^{56}$Ni mass in spite of a high explosion energy of more than 2\,B. The thermal-bomb explosions of the remaining progenitors produce reasonable amounts of nickel. 

For the thermal-bomb explosions with other energy injection timescales ($t_\mathrm{inj}=0.2$\,s and $t_\mathrm{inj}=2.0$\,s) the nickel masses falling back are not significantly different from the results with our default injection time of 1\,s.

\begin{figure*}
	\includegraphics[width=0.99\columnwidth]{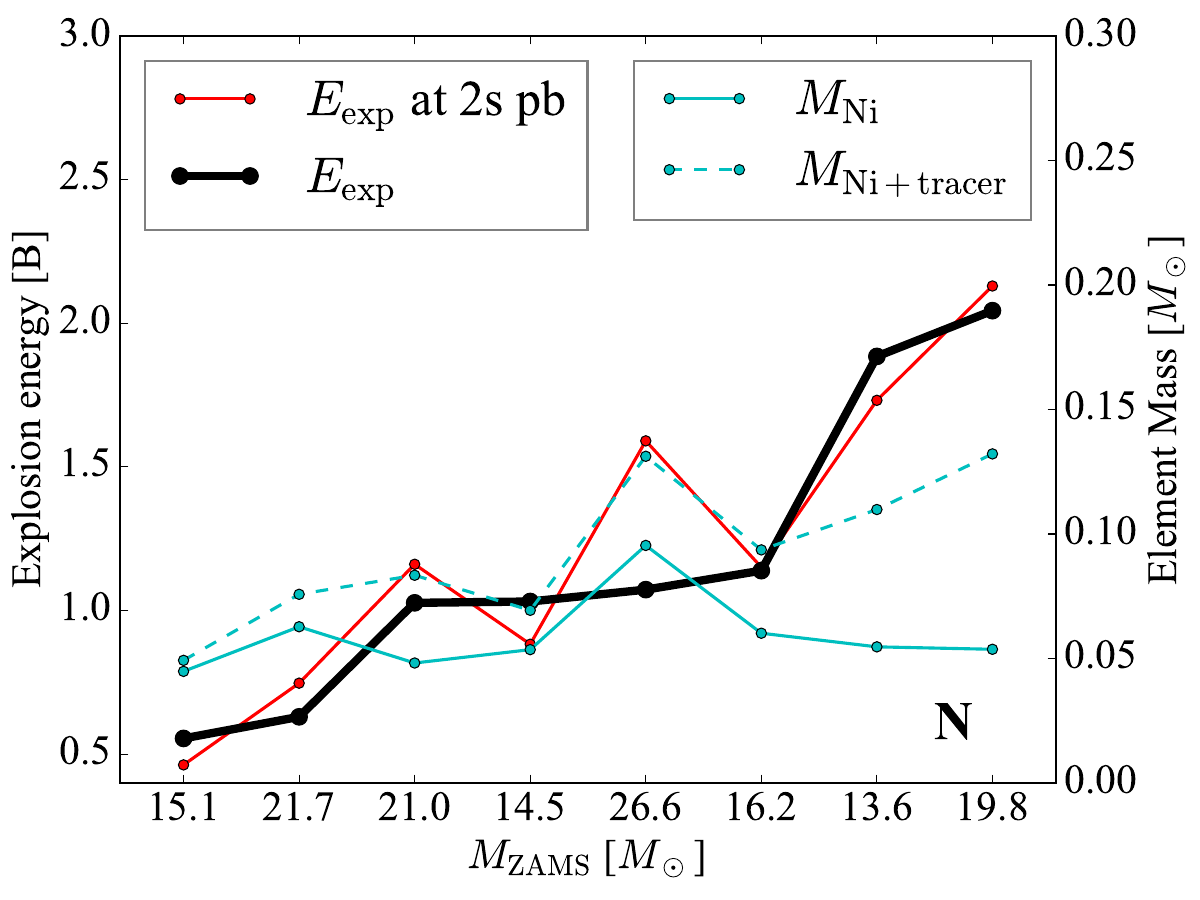}
 	\includegraphics[width=0.99\columnwidth]{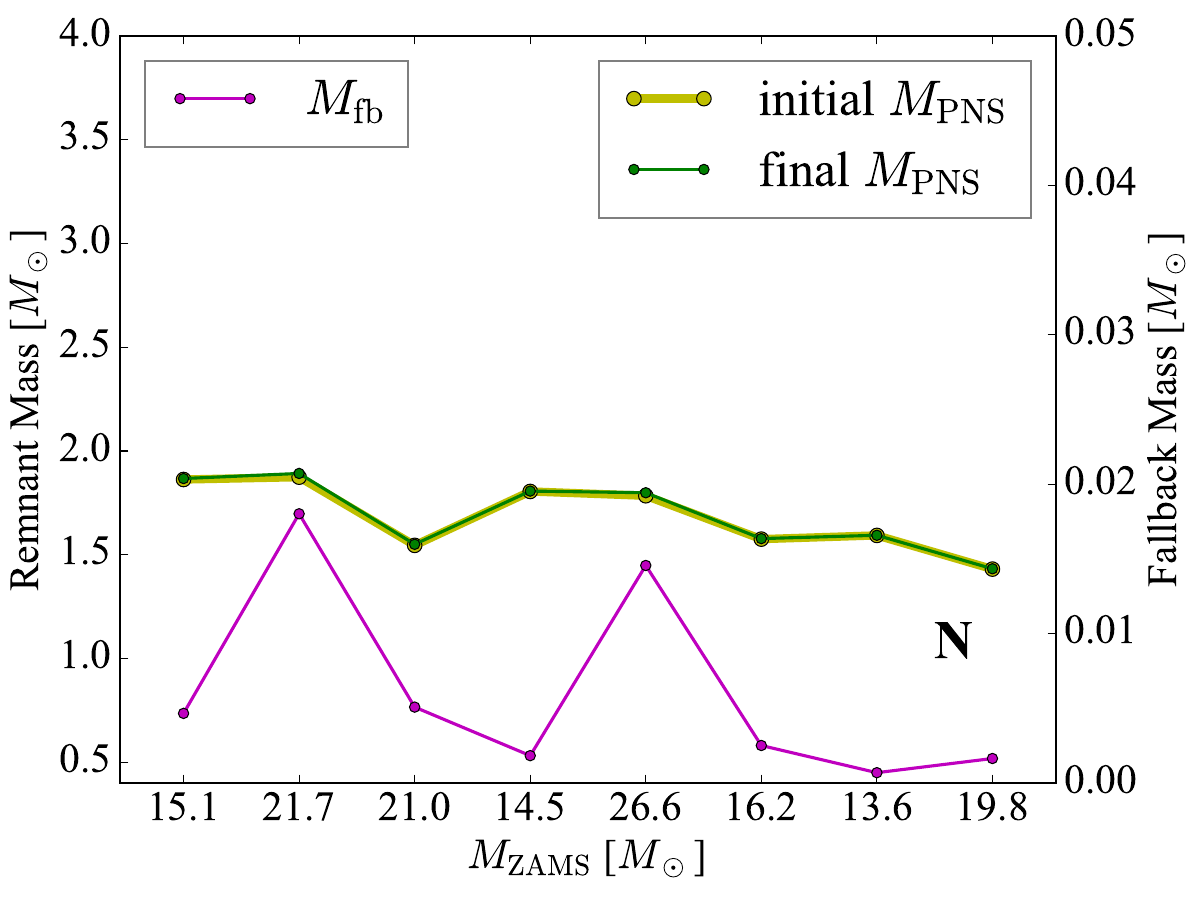}
    \caption{Progenitor dependence of neutrino-driven explosions. Left panel: final explosion energies and explosion energies at 2\,s after bounce (left axis); ejected masses of nickel and of nickel plus tracer (right axis). 
    Right panel: initial and final remnant masses (left axis); fallback masses (right axis). Initial and final remnant masses actually overlap because of the small amount of fallback. The abscissas of both panels are ordered according to increasing explosion energy. The letter ``N'' in the lower right corners indicates that these are the results of the neutrino-driven explosions.}
    \label{fig:summary_neutrino}
\end{figure*}

\begin{figure*}
	\includegraphics[width=0.99\columnwidth]{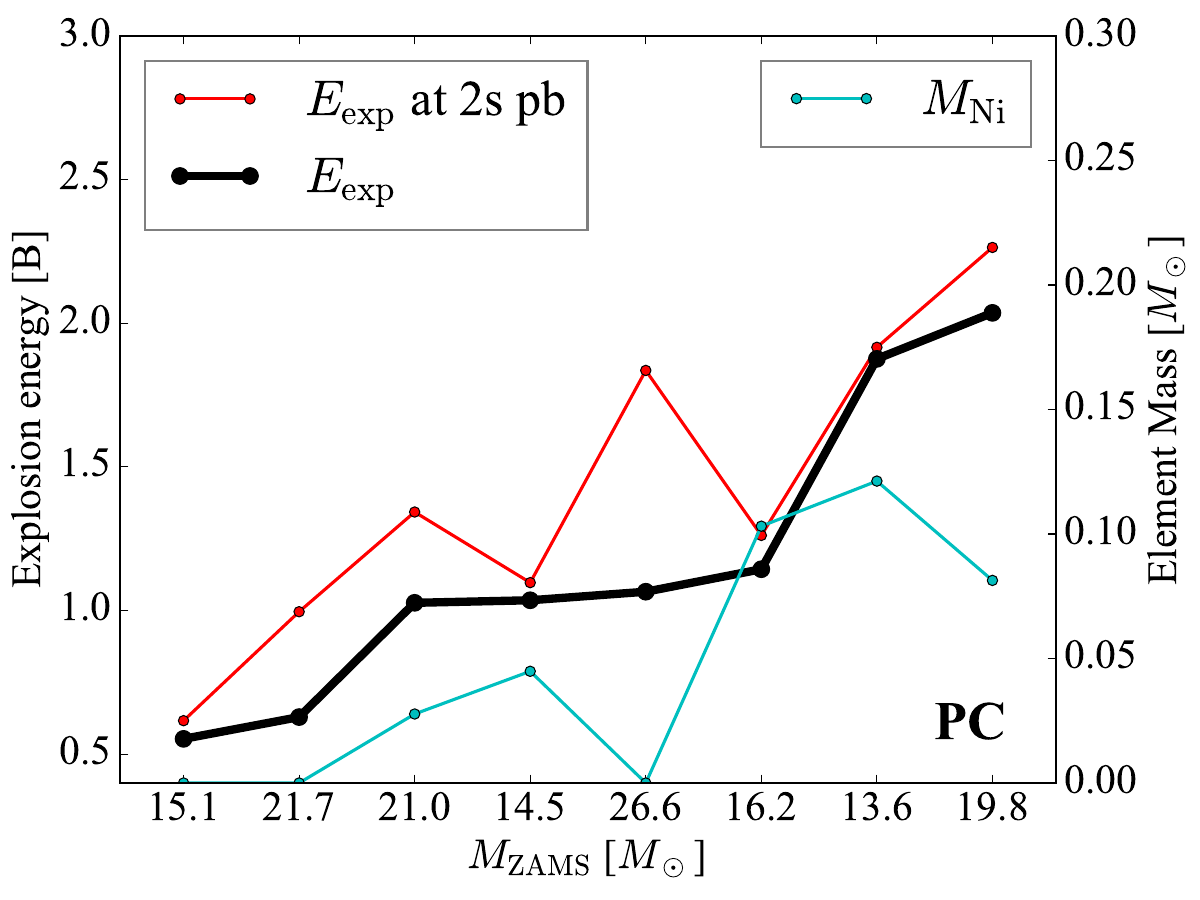}
 	\includegraphics[width=0.99\columnwidth]{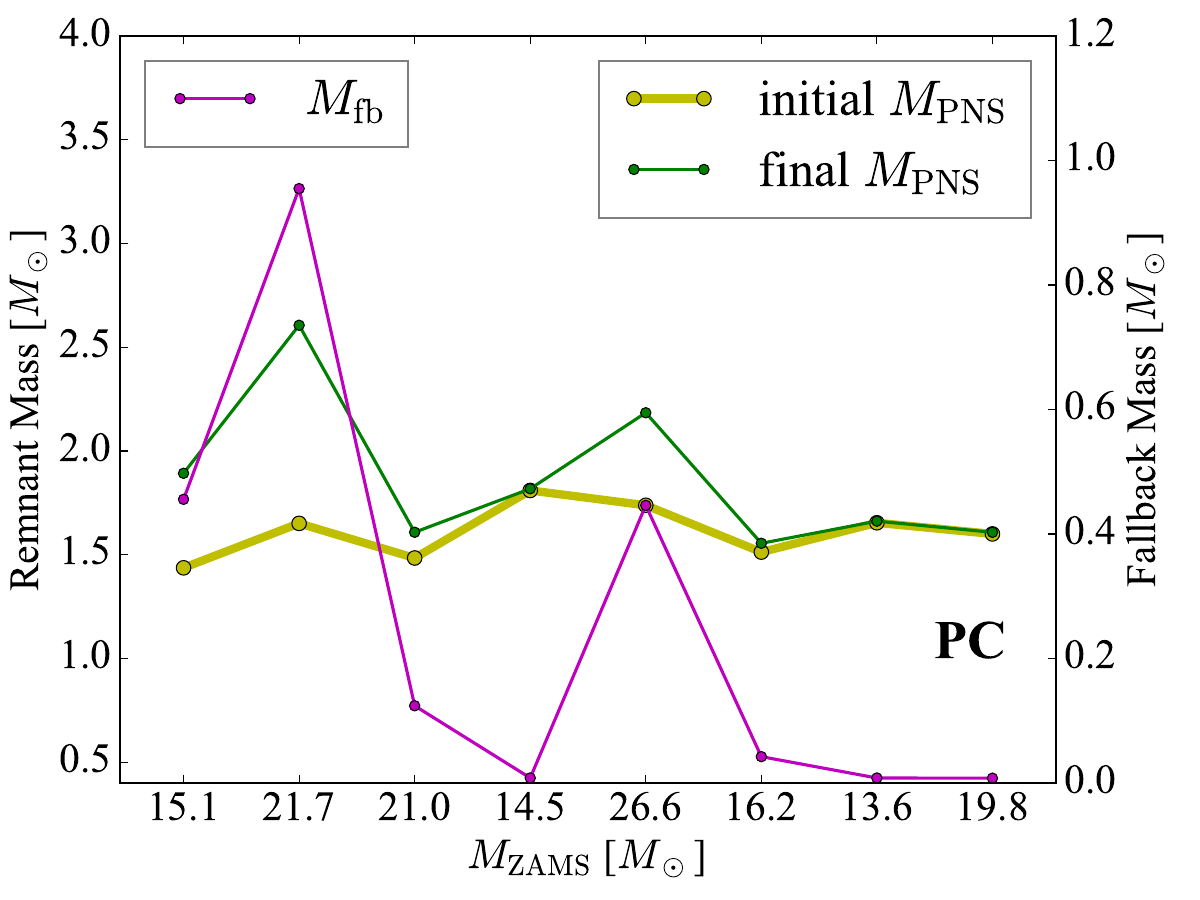}
    \caption{Progenitor dependence of classic piston-driven explosions (PC). Left panel: final explosion energies and explosion energies at 2\,s post bounce (left axis); ejected nickel masses (right axis). 
    Right panel: initial and final remnant masses (left axis); fallback masses (right axis). Abscissas as in Figure~\ref{fig:summary_neutrino}.}
    \label{fig:summary_piston}
\end{figure*}
\begin{figure*}
	\includegraphics[width=0.99\columnwidth]{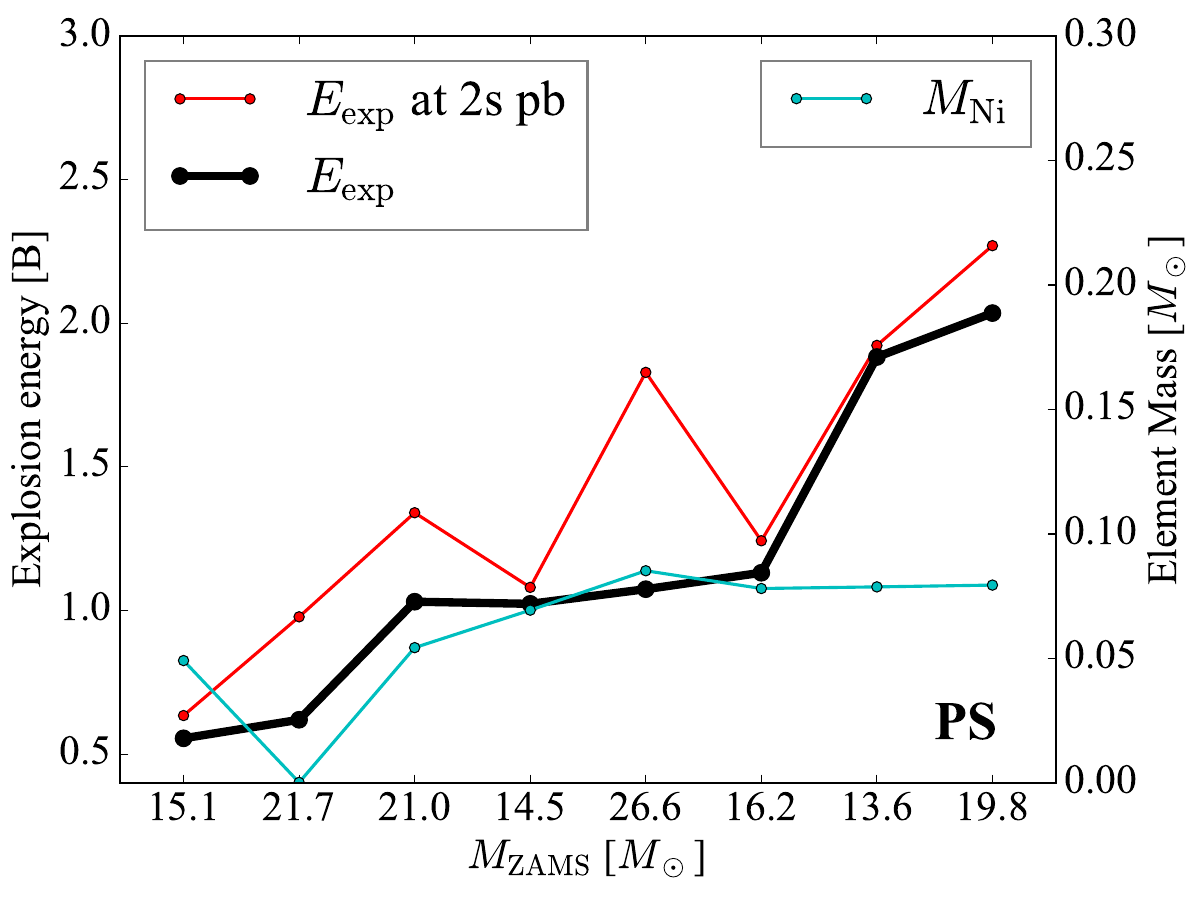}
 	\includegraphics[width=0.99\columnwidth]{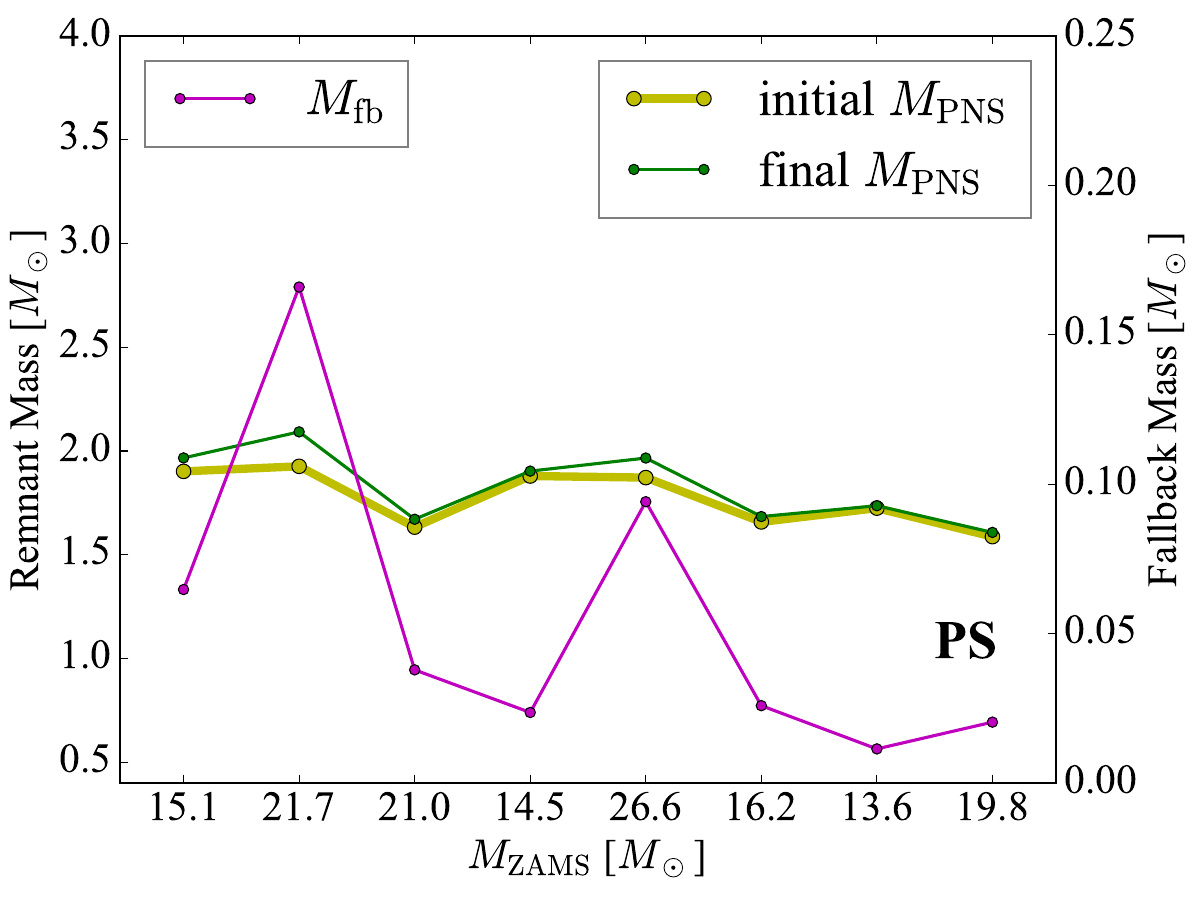}
    \caption{As Figure~\ref{fig:summary_piston}, but for the piston-driven explosions with the special trajectories (PS).}
    \label{fig:summary_piston_special}
\end{figure*}

\begin{figure*}
	\includegraphics[width=0.99\columnwidth]{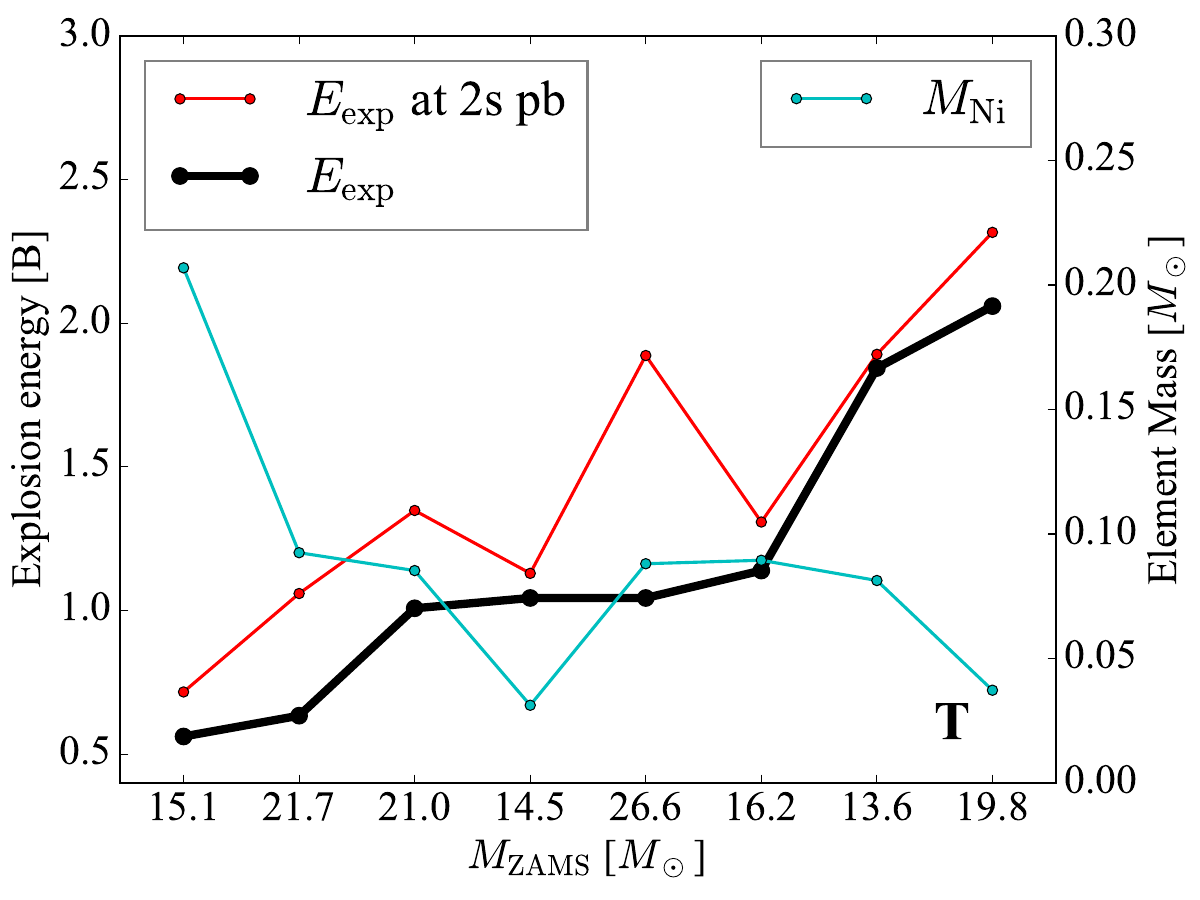}
 	\includegraphics[width=0.99\columnwidth]{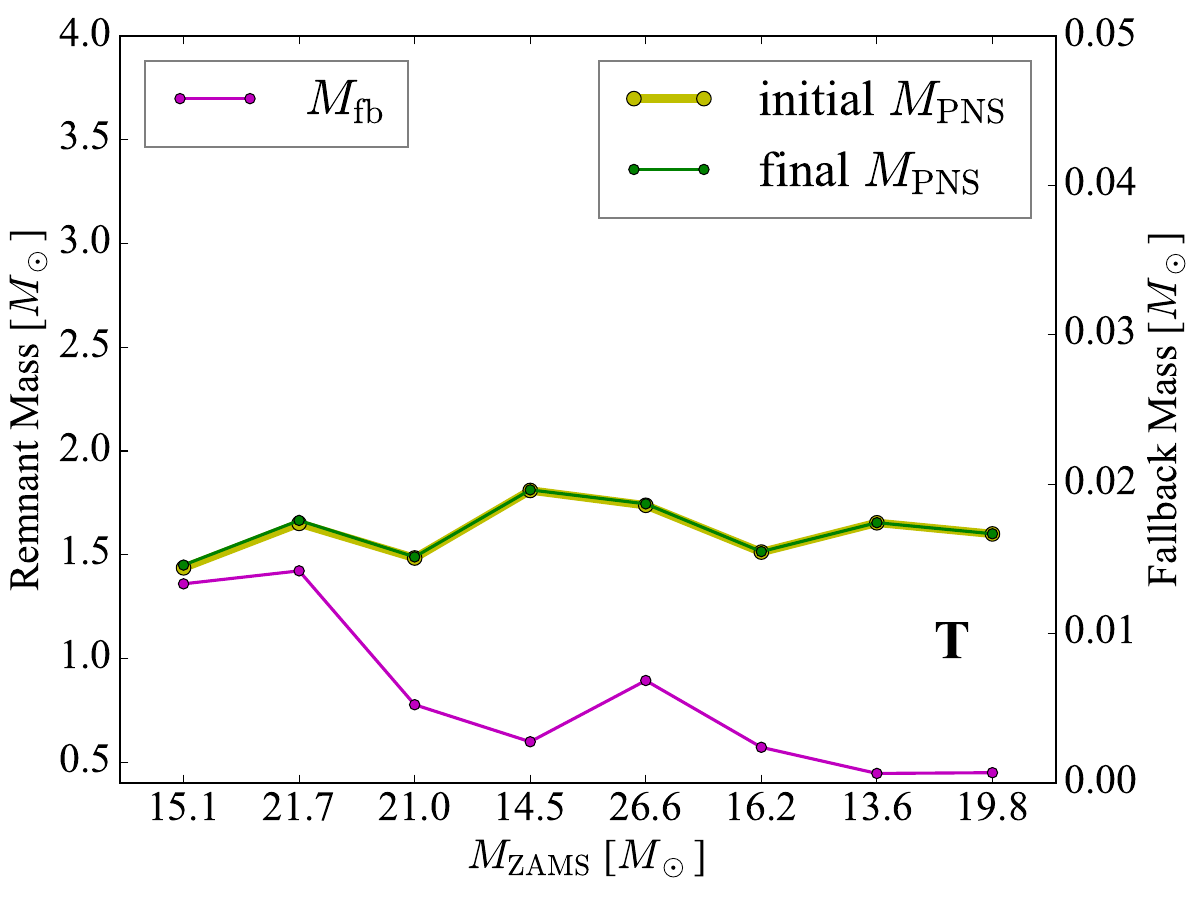}
    \caption{As Figure~\ref{fig:summary_piston}, but for the thermal-bomb explosions (T). Initial and final remnant masses (right panel) overlap again because of the small fallback masses.}
    \label{fig:summary_bomb}
\end{figure*}

\section{Explosion trends across progenitors}
\label{sec:discussion}

The progenitor dependence of the results of all CCSN simulations done in this work is presented in overview in Figure~\ref{fig:summary_neutrino} for neutrino-driven explosions, Figure~\ref{fig:summary_piston} for classical-trajectory piston-driven explosions, Figure~\ref{fig:summary_piston_special} for special-trajectory piston-driven explosions, and Figure~\ref{fig:summary_bomb} for thermal-bomb explosions. In the left panel of each figure we display the final explosion energies (black line), the explosion energies at 2\,s after bounce (red line), the ejected masses of nickel (cyan line), and of nickel plus tracer nucleus (cyan dashed line; for the neutrino-driven case only). The right panels show the initial (yellow line) and final (green line) PNS masses, and the fallback masses (magenta line; note that the scales for the fallback masses on the right ordinates are different for the different explosion mechanisms). For neutrino-driven and thermal-bomb explosions the initial and final PNS masses actually overlap due to the very low fallback masses.  

The results in Figures~\ref{fig:summary_neutrino}--\ref{fig:summary_bomb} are sorted such that the final explosion energy (black lines in the left panels) increases from left to right, with the horizontal axis showing the values of the ZAMS masses, $M_{\rm ZAMS}$, of the corresponding progenitors. Since all explosion models for a given progenitor were tuned to the same explosion energy (as obtained in the neutrino-engine model) within 3\%, the black line is essentially the same for all mechanisms. We also show the explosion energies at 2\,s post bounce (red lines), since this time is close to the phase when the production of iron-group isotopes including nickel takes place. 

For all mechanisms the values of the explosion energies at 2\,s differ from those at the end of the simulations. The general behaviour is quite similar: usually the explosion energy at 2\,s is higher than the final value, because later on the CCSN shock sweeps up overlying progenitor layers with negative binding energy. This order of the two explosion-energy values is inverse in some neutrino-engine models, where the long-lasting energy input by neutrino heating dominates the gravitational binding energy of the progenitor exterior to the shock at 2\,s. Because of the continuous energy input by neutrinos over up to 10 seconds, which (partly) compensates for the binding energy of the overlying progenitor, the differences between the two explosion energy values are smaller in most of the P-HOTB models than in the models with the piston and thermal-bomb mechanisms.

Moreover, for the neutrino-driven explosions (Figure~\ref{fig:summary_neutrino}, left panel), there is a clear correlation between $M_{\rm Ni+tracer}$ (cyan dashed line) and the explosion energy at 2\,s (red line): the higher the energy is, the higher the final mass of nickel plus tracer is, both having peaks for progenitors of 21.0, 26.6, and 19.8\,$M_\odot$. This correlation can be understood by the fact that the energy of neutrino-driven explosions is provided by energy deposition of neutrinos in a part of the ejecta that in the subsequent freeze-out becomes iron-group (and trans-iron) elements, including nickel and tracer material, and, possibly, a remaining mass fraction of $\alpha$-particles (helium). We repeat that some of the tracer material could well be $^{56}$Ni and possibly even a part of the $\alpha$-particles not recombining in the neutrino-heated ejecta of 1D models could well form $^{56}$Ni in 3D explosions of similar energies (see the discussion in Sect.~\ref{sec:niprod} and references given there).\footnote{In this context we point to 3D CCSN results by \citet{Burrows+2024}. In their Fig.~13, left panel, the $^{56}$Ni masses correlate with explosion energies. In the energy range from 0.5\,B to 2.0\,B, which is relevant for our study, this correlation displays considerable scatter with roughly the same $^{56}$Ni mass occurring for different explosion energies. The $^{56}$Ni masses are between $\sim$0.06\,$M_\odot$ and $\sim$0.17\,$M_\odot$. This is not too different from the $^{56}$Ni plus tracer masses of our neutrino-engine models, which range between $\sim$0.05\,$M_\odot$ and $\sim$0.13\,$M_\odot$ with a correlation that also shows some scatter, i.e., similar $^{56}$Ni masses are possible for different explosion energies (see Figure~\ref{fig:summary_neutrino}, left panel).}

A positive correlation between nickel masses and explosion energies is not present in the classic piston-driven and thermal-bomb models. However, there is an anti-correlation between the final nickel masses and the fallback masses (as well as final PNS masses) in the classic piston-driven explosions (Figure~\ref{fig:summary_piston}, compare left and right panels): the higher the fallback mass is, the less nickel manages to get expelled, because most of the nickel is produced close to the PNS, i.e., close to the inner grid boundary. As soon as the inner boundary is opened, nickel is among the first elements to fall back onto the compact remnant.

In neutrino-driven as well as thermal-bomb explosions, the amount of fallback is much smaller than in the two variants of the piston-driven explosions, and in fact the fallback masses are so small that they barely change the final PNS masses (right panels of Figures~\ref{fig:summary_neutrino} and~\ref{fig:summary_bomb}). The variation pattern of the fallback masses with the progenitors (magenta lines) is similar for all of the explosion mechanisms. There are always peaks for the progenitors of 21.7\,$M_\odot$ (low explosion energy) and 26.6\,$M_\odot$ (high binding energy), and the fallback is always lowest for the progenitors with the highest explosion energies ($M_\mathrm{ZAMS}=13.6$\,$M_\odot$ and 19.8\,$M_\odot$). The final masses of the compact remnants of all progenitors are somewhat higher for the special-trajectory piston-driven explosions than for the neutrino-driven explosions, where the neutrino-driven wind reduces the PNS mass (as explained in Section~\ref{sec:fallphase}), and also compared to the thermal-bomb explosions, where the remnant masses tend to be even lower in most cases than in the P-HOTB simulations with the neutrino engine.

It is not very surprising that the $^{56}$Ni masses in the neutrino-driven explosions are, overall, best reproduced by the piston-driven models with special trajectories (Figure~\ref{fig:summary_piston_special}, left panel). This also holds true for the variation of the fallback masses with the progenitors' ZAMS masses (Figure~\ref{fig:summary_piston_special}, right panel), although the values of the fallback masses are still roughly a factor of 10 higher in the special-trajectory piston-driven explosions compared to the neutrino-driven explosions. 

The $21.7M_\odot$ explosion model is an exception. Despite producing a high nickel mass with the special trajectory ($M_{\rm Ni}\sim 0.12\,M_\odot$; Figure~\ref{fig:ni}, lower left panel), this model is the only one with special-trajectory pistons where all of the nickel is removed from the ejecta by fallback, as mentioned in Section~\ref{sec:niprod}. This can again be understood by considering the two main aspects that determine the fallback, namely the explosion energy and the progenitor structure. The progenitors of $21.7M_\odot$ and $15.1M_\odot$, for example, have very similar special-trajectory-piston setups, because their inner grid boundaries (i.e., piston locations) are placed at nearly the same initial mass cuts (1.93\,$M_\odot$ and 1.90\,$M_\odot$ respectively; Table~\ref{tab:piston_special_res}), and the explosion energies in both cases are fairly low and similar (0.62\,B and 0.56\,B, respectively; Table~\ref{tab:piston_special_res}). However, looking at the density structure (Figure~\ref{fig:psn_closer}, upper right panel), one can see that the 21.7\,$M_\odot$ progenitor displays a higher density in the region close to the inner boundary (around 1.9\,$M_\odot$). This ultimately leads to higher fallback, for which reason all of the nickel, which was produced close to the inner boundary, fails to be ejected, in contrast to the 15.1\,$M_\odot$ case, where the special-trajectory explosion yields 0.05\,$M_\odot$ of ejected $^{56}$Ni (Table~\ref{tab:piston_special_res} and lower left panels of Figures~\ref{fig:ni} and~\ref{fig:summary_piston_special}). A dramatic impact of massive fallback, however, is visible also in the 15.1\,$M_\odot$ explosion with the classic piston, where the inner grid boundary is located much deeper inside the progenitor at a mass of only 1.44\,$M_\odot$ (the smallest initial PNS mass in Table~\ref{tab:piston_res}) in a high-density region (Figure~\ref{fig:psn_closer}, upper right panel). Despite the correspondingly high production of $^{56}$Ni (the highest value of all classic piston-driven explosions; Figure~\ref{fig:ni}, upper right panel), the entire nickel finally falls back because of the combination of a relatively low explosion energy and a relatively high gravitational binding energy (Table~\ref{tab:piston_res}).

As exemplified by the examples above, the behavior of individual (classic and special-trajectory) piston-driven and thermal-bomb models, in particular also of outliers and special cases, can be explained by the crucial factors that determine the $^{56}$Ni yields, fallback masses, and remnant masses, namely the explosion energy and the progenitor structure. As for the latter, the position of the inner grid boundary has a sensitive influence through the matter density around this mass cut and, closely linked to it, through the gravitational binding energy of the progenitor outside this point. For a given value of the explosion energy, a high density at the mass cut, which is usually correlated with a high binding energy and a location deep inside the progenitor, does not only facilitate a high production of nickel but also a large fallback mass, in particular for low explosion energies. The lower the explosion energy is, the higher the fallback will be, since there is not enough energy to push all overlying matter sufficiently strongly to make it gravitationally unbound. Conversely, if the explosion energy is relatively high (as in the 13.6\,$M_\odot$ explosion with $\sim$1.9\,B) and/or if the density (binding energy) at the initial mass cut is low (as in the 14.5\,$M_\odot$ model with a mass cut at $\sim$1.81\,$M_\odot$ for the classic piston and thermal bomb) the fallback mass tends to be low. Finally, although a high explosion energy favors a large production of nickel and little fallback, the ultimate yield of $^{56}$Ni can be small if a (relatively) high explosion energy is combined with a very low density around the initial mass cut. This situation exists, for example, for the 14.5\,$M_\odot$ explosions with the classic piston and thermal-bomb mechanisms, where the fallback is small but also the ejected $^{56}$Ni mass is low (0.03--0.045\,$M_\odot$) in spite of an explosion energy of more than 1\,B. This low nickel production can be cured ($M_\mathrm{Ni}\sim 0.07\,M_\odot$) by the special trajectory (despite 3--9 times higher fallback masses), because the matter around the inner grid boundary is first collapsed deeper into the gravitational potential of the forming PNS and thus compressed, before it is shock-heated and expelled. In contrast, even the special trajectory cannot cure the grossly overestimated fallback in the 21.7\,$M_\odot$ explosion with the classic piston compared to the neutrino-driven explosion, which does not permit any or only a tiny amount of $^{56}$Ni to be ejected in both piston-driven explosions. In this case the thermal-bomb mechanism is doing better with resonable amounts of fallback and ejected nickel (Tables~\ref{tab:neutrino_res}--\ref{tab:piston_special_res}). 

Completing the cross-comparison at this point, we summarize that the dependencies of the explosion results on the piston and thermal-bomb setups (mass cut location and progenitor properties) discussed above can explain the general trends and the progenitor variations of the fallback masses, remnant masses, and nickel yields with the explosion energies displayed in Figures~\ref{fig:summary_neutrino}--\ref{fig:summary_bomb} and listed in Tables~\ref{tab:neutrino_res}--\ref{tab:piston_special_res}. In particular, one can see local maxima of the fallback mass always for the same progenitors, because the combination of low/high explosion energy and low/high binding energy of a given progenitor tends to be robust, independent of the explosion mechanism and thus the location of the initial mass cut. Changing the mechanism therefore has an impact on the amount of fallback, as discussed above, but not on the progenitor-dependent pattern, since the explosion energy is kept fixed for each progenitor. Interestingly, placing the mass cut at a less deep position as it is usually the case in the special-trajectory piston models compared to the classic piston-driven explosions (except in the 19.8\,$M_\odot$ simulations, where it is insignificantly reverse), does not necessarily imply less fallback as one might expect because of a lower progenitor density around the inner grid boundary. The 13.6, 14.5, and 19.8\,$M_\odot$ simulations with the special-trajectory piston have slightly more fallback, because the special trajectories are contracted to a smaller radius at bounce, which influences the nickel production as well as the fallback mass.

\section{Summary and conclusions}
\label{sec:conclusion}

In our study we investigated different numerical methods to trigger CCSN explosions in 1D simulations. Considering eight solar-metallicity red supergiant progenitors with ZAMS masses between 13.6\,$M_\odot$ and 26.6\,$M_\odot$, we compared neutrino-driven explosions employing the neutrino engine in the P-HOTB code \citep[progenitors and explosion models from][]{2016ApJ...821...38S} with (a) the classic piston mechanism, (b) a variant of it that uses a specially chosen mass trajectory from the neutrino-engine models, and (c) thermal-bomb explosions with an energy deposition timescale $t_{\rm inj}=1.0$\,s. The classic piston-driven and thermal-bomb explosions were computed with an initial mass cut (i.e., a Lagrangian inner grid boundary) at the $s/k_\mathrm{B} = 4$ location of the pre-collapse stars, following previous literature. The explosion energies, however, were tuned to the values obtained for each individual progenitor in the neutrino-driven explosions. In the special-trajectory models the location of the initial mass cut was taken from the special trajectory, which also defined, progenitor-specific, a new collapse timescale and minimum radius for the bounce at the end of the collapse phase. Our results for the $^{56}$Ni yields, fallback masses, and initial and final PNS masses, ordered in sequence with rising explosion energy, are presented in overview in Figures~\ref{fig:summary_neutrino}--\ref{fig:summary_bomb}. Quantitative information is collected in Tables~\ref{tab:neutrino_res}--\ref{tab:piston_special_res}. 

Our goal was to assess whether simple trigger-recipes for explosions in 1D, which are still widely used to compute large sets of CCSN models, are able to reproduce crucial explosion properties of neutrino-driven 1D explosions, with a special focus on the masses of the compact remnants and of the ejected $^{56}$Ni as a representative, diagnostically important iron-group nucleus. The comparison was eased and constrained by adopting the explosion energies from the neutrino-engine models also for the other three explosion-trigger methods (a)--(c).

Overall, we found that these other methods have a hard time to reproduce the basic properties of interest as obtained in the neutrino-driven explosions. One of the most important consequences of the latter is a correlation of explosion energy and $^{56}$Ni mass (mass of $^{56}$Ni plus neutron-rich tracer nucleus in the P-HOTB models), which is a generic outcome of neutrino-driven explosions in 1D \citep{2016ApJ...821...38S,2020ApJ...890...51E} as well as multi-D \citep{Nakamura+2015,Janka2017,Burrows+2024}. This correlation is not witnessed in the classic piston-driven explosions, neither without fallback nor with fallback, which is massively overestimated by the classic piston method and carries back all initially produced $^{56}$Ni in several of our simulated cases, even though the inner grid boundary is opened and fallback sets in only at 100\,s after bounce. The massive fallback also leads to grossly overestimated final compact-remnant masses in the classic piston-driven explosions.

This problem can be cured to a large extent by adopting the special trajectory for the piston method as applied by \citet{2016ApJ...821...38S}. The fallback masses are considerably reduced in all cases, the ejected nickel masses after fallback are close to those in the neutrino-driven explosions (except in the 21.7\,$M_\odot$ model, which is an outlier with the highest fallback), and, in particular, the correlation of explosion energy and $^{56}$Ni ($^{56}$Ni+tracer, respectively) mass is reproduced. The fallback masses still exceed those in the neutrino-engine models by roughly a factor of 10, and the final compact-remnant masses are also somewhat higher (by up to $\sim$0.1\,$M_\odot$), but the overall trends with progenitor and explosion energy of the P-HOTB models are reproduced.

The thermal-bomb models with the $s/k_\mathrm{B} = 4$ location for the initial mass cut give a mixed picture. Although their fallback masses are similarly low as in the neutrino-driven models, despite the inner grid boundary being opened already at 10\,s after bounce, the PNS masses tend to be lower for many, though not all, cases, because the chosen boundary location is incompatible with the explosion dynamics of the neutrino-engine models. Therefore the thermal-bomb explosions do not display the trend of a mild anti-correlation of the compact-remnant masses with the explosion energies seen in the P-HOTB models. Moreover, they awfully fail to yield the correlation between explosion energy and $^{56}$Ni ($^{56}$Ni+tracer, respectively) mass. It is interesting to note that classic piston explosions and thermal-bomb explosions show basically the same order of the initially produced $^{56}$Ni masses (before fallback) with the progenitors, however the thermal bombs typically create only about 40\%--80\% of the $^{56}$Ni yields of the piston models. Since the fallback is small in the thermal-bomb simulations, it basically does not change this order of the produced nickel masses, whereas the massive fallback in many classic piston cases drastically changes the ultimately ejected $^{56}$Ni masses.    

\citet{1991ApJ...370..630A} found smaller differences in the nickel production between thermal bombs and piston-driven explosions than obtained in our models before fallback, which these authors did not take into account in contrast to our study. The discrepancy of the results is probably connected to differences in the setups of the explosion-trigger mechanisms, which we did not tune for best agreement of the results. In contrast to \citet{1991ApJ...370..630A}, \citet{2007ApJ...664.1033Y} reported more substantial differences in the nickel production of thermal bombs and piston-driven explosions with about three times higher yields for the pistons. Their result therefore reflects a trend similar to what we also witnessed in our simulations before fallback, despite the fact that \citet{2007ApJ...664.1033Y} applied a significantly different modeling approach, simulating the core collapse with neutrino transport, which was switched off after bounce to be replaced by a piston or thermal bomb. Also the inner grid boundary was handled in a significantly different way as absorbing boundary right after the energy injection. 

Despite the overall similarity of our findings and the results reported for piston-driven and thermal-bomb explosions by \citet{1991ApJ...370..630A} and \citet{2007ApJ...664.1033Y}, quantitative and qualitative aspects do not agree because of relevant differences in the modelling setups. From our study one may conclude that special-trajectory pistons are a reasonably good alternative to employing neutrino engines for triggering CCSN explosions in 1D simulations. However, this approach requires the availability of existing calculations of neutrino-driven explosions for each individual investigated case to define the explosion energy and to set the piston parameters via a chosen special trajectory. Thermal bombs perform somewhat better than the classic piston mechanism in various aspects, but they also have grave shortcomings, as we showed in our study. Future users of these methods, which will certainly remain a valuable tool for computationally inexpensive CCSN calculations in 1D (or multi-D) for large grids of progenitor models, should be aware of these limitations.

Although we referred to results from 1D CCSN models with a neutrino engine to judge the performance of the other investigated explosion triggers, we stress that neutrino-driven explosions in 1D cannot capture crucial physics of CCSNe, which are a generically multi-dimensional phenomenon involving neutrinos, turbulent processes, large-scale asymmetries, mixing processes, and varying thermodynamic and nucleosynthetic conditions in multi-component flows. 1D neutrino-driven explosions are therefore unable to describe the element and isotope formation that takes place in the innermost, neutrino-heated CCSN ejecta \citep{Wanajo+2011,2016ApJ...818..123B,Wanajo+2018,Sieverding+2023,Wang+2024a,Wang+2024b}. Moreover, although they yield the positive correlation between explosion energy and nickel mass, which is absent in classic piston and thermal-bomb explosions, they display other trends that disagree with multi-dimensional results, at least for the parameter settings employed in our current P-HOTB neutrino-engine treatment. For example, the mild anti-correlation of PNS mass and explosion energy seen in the P-HOTB results is in conflict with the strong, positive correlation obtained in self-consistent 2D and 3D simulations \citep{Nakamura+2015,Burrows+2024}, which can be understood by fundamental differences of the explosion dynamics and accretion and outflow (neutrino-driven wind) history of the new-born PNS in 1D P-HOTB and multi-D simulations.

\section*{Acknowledgements}
\label{sec:thanks}

We extend our appreciation to Thomas Ertl for his support during the project's initial phase, and we express our gratitude to Ewald M\"uller and Johannes Ringler for their valuable discussions. 
Support by the German Research Foundation (DFG) through the Collaborative Research Centre ``Neutrinos and Dark Matter in Astro- and Particle Physics (NDM),'' grant No. SFB-1258-283604770, and under Germany's Excellence Strategy through the Cluster of Excellence ORIGINS EXC-2094-390783311 is acknowledged.

\section*{Data Availability}
\label{sec:data}

The data of our calculations will be made available upon reasonable request.

\section*{Software}
\textsc{Prometheus-HOTB}  \citep{1996A&A...306..167J,2003A&A...408..621K,2006A&A...457..963S,2007A&A...467.1227A,2012ApJ...757...69U,2016ApJ...818..124E};
KEPLER \citep{1978ApJ...225.1021W}; Matplotlib \citep{2007CSE.....9...90H}; NumPy \citep{2011CSE....13b..22V}.



\bibliographystyle{mnras}
\bibliography{imasheva_references} 







\bsp	
\label{lastpage}
\end{document}